\documentclass[aps,preprintnumbers,floatfix,article,amsmath,amssymb,floatfix,10pt,prd,superscriptaddress,nofootinbib]{revtex4}
\usepackage{bm}
\usepackage{amsfonts}
\usepackage{latexsym}
\usepackage[latin1]{inputenc}
\usepackage{graphicx}
\usepackage{amsmath}
\usepackage{palatino}
\usepackage{mathpazo}
\usepackage{textcomp}
\usepackage{float}
\usepackage{booktabs}
\usepackage{dcolumn}
\usepackage{ragged2e}
\usepackage{hyperref}
\hypersetup{colorlinks,citecolor=blue}
\hypersetup{colorlinks=true,linkcolor=red,filecolor=magenta,    urlcolor=cyan}
\usepackage{amsmath}
\usepackage{xcolor}
\usepackage{epsfig}
\usepackage{caption}
\usepackage{commath}

\def\jnl@style{\it}
\def\aaref@jnl#1{{\jnl@style#1}}

\def\aaref@jnl#1{{\jnl@style#1}}

\def\aj{\aaref@jnl{AJ}}                   
\def\apj{\aaref@jnl{ApJ}}                 
\def\apjl{\aaref@jnl{ApJ}}                
\def\apjs{\aaref@jnl{ApJS}}               
\def\apss{\aaref@jnl{Ap\&SS}}             
\def\aap{\aaref@jnl{A\&A}}                
\def\aapr{\aaref@jnl{A\&A~Rev.}}          
\def\aaps{\aaref@jnl{A\&AS}}              
\def\mnras{\aaref@jnl{Mon.~Not.~Roy.~Astron.~Soc.}}             
\def\prd{\aaref@jnl{Phys.~Rev.~D}}        
\def\prc{\aaref@jnl{Phys.~Rev.~C}}  
\def\prl{\aaref@jnl{Phys.~Rev.~Lett.}}    
\def\qjras{\aaref@jnl{QJRAS}}             
\def\skytel{\aaref@jnl{S\&T}}             
\def\ssr{\aaref@jnl{Space~Sci.~Rev.}}     
\def\zap{\aaref@jnl{ZAp}}                 
\def\nat{\aaref@jnl{Nature}}              
\def\aplett{\aaref@jnl{Astrophys.~Lett.}} 
\def\apspr{\aaref@jnl{Astrophys.~Space~Phys.~Res.}} 
\def\physrep{\aaref@jnl{Phys.~Rep.}}      
\def\physscr{\aaref@jnl{Phys.~Scr}}       
\def\commat{\aaref@jnl{Comm.~Math.~Phys.}}              
\def\science{\aaref@jnl{Science}}               
\def\cqg{\aaref@jnl{Classical Quant.~Grav.}}            
\def\jpcs{\aaref@jnl{JPCS}}                                     
\def\ijmpd{\aaref@jnl{Int.~J.~Mod.~Phys.~D}}                    
\def\grg{\aaref@jnl{Gen.~Relat.~Gravit.}}               
\def\rpp{\aaref@jnl{Rep.~Prog.~Phys.}}          
\def\npa{\aaref@jnl{Nucl.~Phys.~A}}        
\def\lrr{\aaref@jnl{Living Rev.~Rel.}}                   
\def\jcap{\aaref@jnl{J.~Cosmology Astropart.~Phys.}}    
\def\rmp{\aaref@jnl{Rev.~Mod.~Phys.}}   
\def\epjc{\aaref@jnl{Eur.~Phys.~J.~C}} 
\def\plb{\aaref@jnl{~Phy.~Lett.~B}} 
\def\mpla{\aaref@jnl{Mod.~Phy.~Lett.~A}} 
\def\arxiv{\aaref@jnl{arxiv.org}}

\allowdisplaybreaks[1]

\begin{document}
\title{\bf Traversable wormhole models in $f(R)$ gravity}

\author{B. Mishra}
\email{bivu@hyderabad.bits-pilani.ac.in}
\affiliation{Department of Mathematics,
Birla Institute of Technology and Science-Pilani, Hyderabad Campus,
Hyderabad-500078, India.}

\author{A. S. Agrawal}
\email{agrawalamar61@gmail.com}
\affiliation{Department of Mathematics,
Birla Institute of Technology and Science-Pilani, Hyderabad Campus,
Hyderabad-500078, India.}

\author{S. K. Tripathy}
\email{tripathy\_sunil@rediffmail.com}
\affiliation{Department of Physics, Indira Gandhi Institute of Technology, Sarang, Dhenkanal, Odisha 759146, India}

\author{Saibal Ray}
\email{saibal@associates.iucaa.in}
\affiliation{Department of Physics, Government College of Engineering and Ceramic Technology, Kolkata 700010, West Bengal, India}

\begin{abstract}
In this work, we analyze the wormhole solutions in $f(R)$ gravity. Specifically we sought for wormhole geometry solutions for the following three shape functions: (i) $b(r)=r_{0}+\rho_{0}r_{0}^{3}\ln\left(\frac{r_{0}}{r}\right)$, (ii) $b(r)=r_{0}+\gamma r_{0}\left(1-\dfrac{r_{0}}{r}\right)$, and (iii) $b(r)=\alpha +\beta r$, under some legitimate physical conditions on the parameters as well as constants involved here with in the shape functions. It is observed from the graphical plots that the behaviour of the physical parameters are interesting and viable.\\

{\bf keywords}: Traversable wormhole; $f(R)$ gravity; perfect fluid; anisotropic fluid
\end{abstract}

\maketitle

\section{Introduction}
The possibility of traversable wormhole in General Relativity (GR) has been independently and simultaneously  demonstrated by Ellis \cite{Ellis73} and Bronnikov~\cite{Bronnikov73}. The Ellis drainhole has been regarded as the first complete model of such a wormhole.  Unaware of Ellis wormhole, Morris and Thorne~\cite{Morris88} proposed the traversable wormhole which was held by a spherical shell of exotic matter. This is popularly known as Morris-Thorne wormhole. They have prescribed  nine properties; some of them are mandatory for the existence of traversable wormhole. The basic wormhole criteria: the metric should be time independent and spherically symmetric, solution should obey the Einstein's field equations, the solution must have a throat to connect and non-existence of horizon.   Later Visser \cite{Visser89} has given some simple examples of traversable wormhole and indicated that the traversing path in the wormhole does not have the region of exotic matter. The traversable   wormhole is the hypothetical tunnels that connects two regions of the universe or two different universes.  The throat of the wormhole is the minimal surface area that linked to satisfy the flare-out condition and exotic matter is the matter that violates null or weak energy condition.  Basically the study of wormhole geometry is to identify an appropriate geometry, which will minimize the exotic matter content at the throat of the wormhole. 

So, the traversable wormholes are hypothetical tunnels in space time given by the following line element~\cite{Morris88}
\begin{equation}\label{eqn.5}
ds^2=-e^{2\Phi(r)}dt^2+\frac{1}{1-b(r)/r}dr^2+r^2(d\theta^2+\sin^2\theta d\phi^2),
\end{equation}
where $(t,~r,~\theta,~\phi )$ are the usual space time spherical coordinates, and $\Phi(r)$ and $b(r)$ are arbitrary functions of the radial coordinate $r$. $\Phi(r)$ determines the gravitational redshift, hence called it as the redshift function, and $b(r)$ determines the spatial shape of the wormhole, hence called the shape function. The wormhole throat corresponds to a minimum value of the radial coordinate, usually denoted by $b_0$ or $r_0$ in the literature. From Eq. \eqref{eqn.5}, we can derive the Ricci scalar as $R=\frac{2b'(r)}{r^2}$, where the notation prime denotes the derivative with respect to the radial coordinate $r$.

The energy conditions are important when we deal with the energy and matter in General Relativity. Also the strong energy condition (SEC) should violate in modified theories of gravity. It has been indicated that the stability of wormhole solution depends on the exotic matter content and exotic matter content should violate the null energy condition (NEC). The NEC along the radial and tangential direction can be represented respectively as, $\rho+p_r$ and $\rho+p_t$. The weak energy condition (WEC) though weaker than strong energy condition and dominant energy condition (DEC) but treated superior to null energy condition can be represented as $\rho >0$ and $\rho +p_r\geq 0$ along radial direction and $\rho >0$ and $\rho +p_t\geq 0$ along tangential direction. 

We wish to make the literature survey on the research on traversable wormhole. Hochberg and Visser~\cite{Hochberg97} have given a geometric structure for generic traversable wormhole.  Krasnikov~\cite{Krasnikov00} have shown that at some stage of the evolution of wormhole, it can generate beam of high energy photons. The traversable wormhole should violate the null energy condition, but the total amount of energy condition violating matter in the space time was not clear until Visser et al.~\cite{Visser03} have developed a suitable approach for its measurement. Lobo~\cite{Lobo05} has pointed out that since phantom energy violates the null energy conditions, this property is required for the sustenance of wormhole. Lobo~\cite{Lobo05} has indicated the expansion and contraction of first and second mouth respectively and because of this a time-shift would be created between both the mouths. Then the wormhole transforms into a machine, and the possible big rip singularity can be avoided. Matos and Nunez~\cite{Matos06} have obtained a solution using Einstein's equation to describe the inner region of a rotating wormhole and observed that slow rotating wormhole stars are more prone to connect a human traveler with more distant regions. 

It is interesting to note that with the numerical techniques and simulation, Gonzalez et al.~\cite{Gonzalez08} have given the non-linear evolution of spherically symmetric wormhole solutions, coupled with massless ghost scalar field and with linear stability analysis complemented that all the wormhole solutions obtained are unstable with respect to linear fluctuations. Rahaman et al.~\cite{Rahaman12} have obtained the wormhole solutions in an extended non-commutative geometry whereas in~\cite{Rahaman15} a new wormhole solution based on non-commutative geometry has been suggested by allowing the conformal Killing vectors. Chew et al. \cite{Chew16} have studied the wormholes solution and analysed the geodesics of these wormholes. They have observed that based on the sign of the asymmetry parameter, the location of the throat shifts towards positive or negative values of constant radial coordinate. In GR, Carvente et al.~\cite{Carvente19} presented the traversable wormhole solutions, that has been supported by a family of ghost scalar fields with quartic potential. It has been realized that the parameter space describing wormhole solutions of the Einstein-scalar field equations enabled to study the properties of exotic matter and their geometry. By embedding class I in GR, Tello-Ortiz and Contreras~\cite{Tello-Ortiz20} have solved the analytical wormhole solutions. 

Post supernovae era, several geometrical modifications have been done in GR to address the late time cosmic acceleration issue. In fact, the universe is believed to have passed through two accelerated expansion phase, the inflationary phase at an early epoch and the dark energy driven accelerated phase at late cosmic times. Usually, within a framework of general relativity, extra fields are considered as a possible approach to understand the dynamics both at the small and large curvature stages of the universe. While GR can provide reasonable explanation at small curvatures, a substantial understanding of the cosmic speed up issue both at small and large curvatures requires a geometrical  modification of GR without the consideration of an extra matter fields \cite{Nojiri11}. One of the most studied geometrically modified gravitational theory is the $f(R)$ theory where we have a more general function of the Ricci scalar $R$ in place of $R$ appearing in the Einstein-Hilbert action \cite{Nojiri2003, Nojiri2007}.  Of late, many viable f(R) cosmological models have been constructed and are used successfully to address issues such as stellar formation and evolution \cite{Astashenok2015, Laurentis2013}, structure formation and evolution \cite{Abdelwahab2008, Hu2015, Terukina2012} and gravitational waves \cite{Clifton2010}. Recently, cosmic objects such as black holes, gravastars, strange stars, and wormholes have been examined using modified gravity theories. The motivation of this paper is to develop  shape function to study of the wormhole solutions in specific form of $f(R)$ gravity, in which $f(R)=\left(\frac{R}{\Upsilon}\right)^{\frac{\alpha}{2}}\left(\frac{2R}{\alpha+2}\right)$ \cite{Tripathy16}. The $f(R)$ Lagrangian cannot be at random. It must adhere to some observational restrictions as well as theoretical mandates. In $f(R)$ gravity, a cosmological model should be stable and capable of simulating a universe that matches observations. In an earlier work \cite{Tripathy16}, the choice of the functional with the $f(R)$ Lagrangian has been shown to be stable and is in conformity with the solar test of Hu and Sawicki \cite{Hu2007}.  The anisotropy in the expansion rates, on the other hand, impacts the functional $f(R)$ via the coefficient ${2}/{[\Upsilon^{\frac{\alpha}{2}}(\alpha +2)]}$. The factor $\Upsilon$ is responsible for the anisotropic effect. Barrow and Clifton \cite{Barrow06} used the formula $f(R)=R^{1+\delta}$ with specified value of $\delta$ in their study. The general formulation of the reconstruction scheme for modified gravity with the $f(R)$ action was presented by Nojiri and Odintsov \cite{Nojiri06}. It is demonstrated how the implicit form of the function $f(R)$ can be defined by any cosmology. The structure and cosmological aspects of traditional $f(R)$ and Hoava-Lifshitz $f(R)$ gravity, scalar-tensor theory, string inspired and Gauss-Bonnet theory, non-local gravity, non-minimally coupled models, and power-counting renormalizable covariant gravity are all investigated \cite{Nojiri11}. Different representations of such theories, as well as their interrelationships, are investigated. Some variations of the preceding theories are consistent with local testing and provide a qualitatively adequate unified explanation of inflation during the dark energy epoch, as well as a thorough cosmological reconstruction of several changing gravities. Extended theories of gravity can be thought of as a new paradigm for solving the infrared and ultraviolet scale problems with general relativity. They are a method for addressing conceptual and practical challenges in astrophysics, cosmology, and high energy physics by keeping the undeniably positive outcomes of Einstein's theory \cite{Capozziello11}. It is worth noting that the traversable wormhole geometry in Einstein's GR violates the null energy condition, $\rho >0$, $NEC_{r}:\rho+p_{r}>0$ and $NEC_{t}:\rho+p_{t}>0$, respectively. To ensure the wormhole's traversability in the modified theory of gravity, we are currently examining the behaviour of the energy conditions in the context of modified $f(R)$ gravity.

The traversable wormhole solution has also been studied extensively in these modified theories of gravity.  Pavlovic and Sossich~\cite{Pavlovic15} have analysed different solutions of wormhole in $f(R)$ gravity and claimed that in high curvature regime all viable $f(R)$ models must have the same mathematical form for the violation of WEC. Zubair et al.~\cite{Zubair16} have claimed that the wormhole solution from anisotropic matter in $f(R,T)$ gravity is stable, viable and obtaining the micro wormhole is realistic, however asymptotic flatness condition is incompatible in case of barotropic fluid. Bhattacharya and Chakraborty~\cite{Bhattacharya17} have analysed the energy conditions in wormhole geometry graphically and obtained the violation of null energy condition at large region around the throat. 

In Rastall gravity, Moradpour et al. \cite{Moradpour17} have presented the traversable asymptotically flat wormhole and indicated the possibility of traversibility of wormhole geometry with phantom sources. The wormhole parameters are observed to be affected by the primary conditions and Rastall dimensionless parameter. Elizalde and Khurshudyan~\cite{Elizalde18} have constructed the wormhole solution in $f(R,T)$ gravity by imposing the condition on the radial pressure to admit the equation of state of varying Chaplygin gas and barotropic fluid. In $f(G,T)$ gravity, Sharif and Ikram~\cite{Sharif18} have explored the wormhole solutions in the galactic region. Tefo et al.~\cite{Tefo19} presented the traversable wormhole solution in $f(T)$ gravity and claimed that with diagonal set of tetrads the static wormhole solution is possible only in nondynamical space-time. Garattini~\cite{Garattini19} proposed the static traversable wormhole, which can explore negative energy density and analysed the consequences of the constraint. Elizalde and Khurshudyan \cite{Elizalde19} obtained wormhole model in $f(R,T)$ gravity. Restuccia and Tello-Ortiz~\cite{Restuccia20} have given the wormhole solution in  $f(R)$ gravity model and studied its cosmological properties. The model presented here is ghost free, satisfies the stability under cosmological perturbations and consistency of local gravity tests. Samanta et al.~\cite{Samanta20} have made a comparison of exponential shape function based traversable wormhole solution in GR and modified gravity. By introducing dimensionful parameter in Brans-Dicke theory, Papantonopoulos and Vlachos~\cite{Papantonopoulos20} have obtained the wormhole solutions. Tripathy~\cite{Tripathy21} has studied the Casimir wormhole solution in the extended gravity. Falco et al.~\cite{Falco21} have reconstructed the wormhole solution in curvature based extended gravity and also indicated that the possibility of obtaining wormhole solutions observationally.

In one of our previous investigations~\cite{Mishra21} on wormhole we have considered following three shape functions: (i) $b(r)=r_0+b_1\left(\frac{1}{r}-\frac{1}{r_0}\right)$, (ii) $b(r)=\sqrt{r_0r}$ and (iii) $b(r)=\frac{r_0^2}{r}$ with the meaning of the symbols therein where interesting features were exhibited. Motivated by these results in the present work we are attempting to study wormholes for some other shape functions to check whether our solutions provide reasonable and physical output. The outline of investigation is arranged as follows: in Sect. 2, we have developed the basic formalism and the filed equations in $f(R)$ gravity. In Sect. 3, the wormhole solutions are obtained in $f(R)$ gravity with three shape functions and the behaviour of the solutions are presented. In Sect. 4, the results and discussions on the wormhole solutions are given.

\section{Basic Formalism and Field Equations}
A special type of modified theory of gravity that generalizes Einstein's General Relativity (GR) is the $f(R)$ gravity, $R$ be the Ricci scalar. The $f(R)$ gravity is a family of gravity theories where each one is defined by different function of the Ricci scalar. The most simplified function is when the function equal to the scalar, which reduces to GR. The objective of introducing the arbitrary function in the modified gravity is that there may be some freedom in explaining the structure formation of the universe and late time cosmic acceleration issue. Buchdahl \cite{Buchdahl70} was first to introduce this where he used $\phi$ in place of the arbitrary function $f$. Starobinsky~\cite{Starobinsky80} extensively used this gravity to study the cosmic inflation problem. Several aspects of the astrophysical and cosmological problems are successfully studied in $f(R)$ theory of gravity. We are intending to study in this paper the wormhole geometry in $f(R)$ theory gravity, whose action can be considered as
\begin{equation} \label{eqn.1}
S=\int\left[\frac{f(R)}{2k}+\mathcal{L}_m\right]\sqrt{-g}d^4x,
\end{equation}
where $k=8\pi G$, with $\mathcal{L}_m$ as the matter Lagrangian and $g$ as the determinant of the metric $g_{\mu \nu}$. We have used the natural system of unit such as, the reduced Planck constant $\hbar$, Newtonian gravitational constant $G$ and speed of light in vacuum $c$. Moreover, $G=\hbar=c=1$ and for simplicity we take $k=1$. Now, with the variation of the action with respect to $g^{\mu\nu}$ and incorporating the metric approach, we can obtain the field equations of $f(R)$ gravity as
\begin{equation} \label{eqn.2}
[R_{\mu \nu}-\nabla_{\mu}\nabla_{\nu}+g_{\mu \nu}\square]F-\frac{1}{2}fg_{\mu \nu}=T_{\mu \nu,}
\end{equation}
where $F$ represents the derivative of $f(R)$ with respect to the Ricci scalar $R$. 

We can obtain the contraction of \eqref{eqn.2} as
\begin{equation} \label{eqn.3}
RF+3\square F-2f=T,
\end{equation} 
where $T=g^{\mu \nu}T_{\mu \nu}$ is the trace of stress energy tensor and incorporating the contraction and rearranging of terms~\cite{Lobo09}, the $f(R)$ gravity field equations can be obtained as
\begin{equation} \label{eqn.4}
G_{\mu \nu}=R_{\mu \nu}-\frac{1}{2}Rg_{\mu\nu}=T_{\mu\nu}^{EF}.
\end{equation}

The R.H.S. of Eq. \eqref{eqn.4} represents the stress energy tensor, $T_{\mu\nu}^{EF}$, which is the combination of curvature stress energy tensor, $ T_{\mu \nu}^{(c)}=\frac{1}{F} \left[\nabla_{\mu}\nabla_{\nu}-\frac{1}{4}(RF+\square F+T)g_{\mu \nu}\right]$ and $ \widehat{T}_{\mu \nu}^{(m)}=\frac{{T}_{\mu \nu}^{(m)}}{F}$; ${T}_{\mu \nu}^{(m)}$ be the matter stress energy tensor.

The energy momentum tensor of matter that describes the matter content of the wormhole, can be defined as an anisotropic distribution of matter as
\begin{equation}\label{eqn.6}
T_{\mu \nu}=(\rho+ p_t)u_{\mu} u_{\nu} +(p_r-p_t)x_{\mu}x_{\nu}+p_tg_{\mu \nu},
\end{equation} 
with $x^{\mu}x_{\mu}=1$, $u^{\mu} u_{\mu}=-1$ . Here $\rho$ denotes the energy density of the matter whereas $p_t$ and $p_r$ are respectively the transverse and radial pressures. The trace, $T=[-\rho,p_r,p_t,p_t]$ and we shall propose the model with zero tidal force ($\Phi=0$). So, the field equations of $f(R)$ gravity \eqref{eqn.4} for the Morris-Thorne wormhole metric \eqref{eqn.5} can be derived and arranged as the energy density, radial pressure and traversal pressure as
\begin{eqnarray} \label{eqn.7}
\rho&=&F\frac{b'}{r^2},\\
p_r&=&F''\frac{b-r}{r}+F'\frac{b'r-b}{2r^2}-F\frac{b}{r^3},\\ \label{eqn.8}
p_t&=&F'\frac{b-r}{r^2}+F\frac{b-b'r}{2r^3}.\label{eqn.9}
\end{eqnarray}

\section{Solutions to wormholes in $f(R)$ Gravity}
In this section, we shall incorporate a specific choice of $f(R)$, where the corresponding We will study the wormhole solution with the function $f(R)$ expressed as $f(R)=\left(\frac{R}{\Upsilon}\right)^{\frac{\alpha}{2}}\left(\frac{2R}{\alpha+2}\right)$, where $\Upsilon = 18\left[\frac{1}{3+\alpha}- \frac{k^2+2k+3}{6(k+2)^2}\right]$ such that $F=\left(\frac{R}{\Upsilon}\right)^{\frac{\alpha}{2}}$~\cite{Tripathy16}. Plane symmetric models with anisotropy in the expansion rates are considered. The anisotropy in expansion rates was thought to persist throughout the evolution of the universe. The condition $F(R)<0$ (Violation of non-existence theorem) is evident when either $[R<0 \rightarrow 2b'/r^{2}<0 \rightarrow b'<0, \Upsilon>0]$ or $[R>0 \rightarrow 2b'/r^{2}>0 \rightarrow b'>0, \Upsilon<0]$. The objective of this specific choice of $f(R)$ is that the deceleration parameter is controlled by the value of $\alpha$. Then we can find
\begin{eqnarray}
F&=&\left(\frac{2b'}{\Upsilon r^2}\right)^{\alpha/2}, \nonumber\\
F'&=&\left(\frac{F}{2rb'}\right)\alpha(rb''-2b'), \nonumber \\
F''&=&\frac{F}{2b'r}\left[\frac{\alpha^{2}-2\alpha}{2b'r}(rb''-2b')^{2}+\frac{\alpha}{r^{2}}(r^{3}b'''-4r^{2}b''+6rb')\right] \nonumber. 
\end{eqnarray} 

We are intending to study the wormhole geometry solutions with the following shape functions:
\begin{itemize}
\item $b(r)=r_{0}+\rho_{0}r_{0}^{3}\ln\left(\frac{r_{0}}{r}\right)$ \cite{Rahaman16} ,
\item $b(r)=r_{0}+\gamma r_{0}\left(1-\dfrac{r_{0}}{r}\right)$ \cite{Rahaman16},
\item $b(r)=\alpha +\beta r$ \cite{Cataldo17},
\end{itemize}
where the shape function $b(r)$ should obey the following conditions:\\
(i) $b'(r_0)\leq1$, (ii) $b(r)<r$, for $r>r_0$ (iii) $\dfrac{b(r)}{r}\rightarrow0$ as $r\rightarrow\infty$, (iv) $\dfrac{b(r)-b'(r)r}{2b^{2}(r)}>0$, $b(r_{0})=r_{0}$; and (v) in order to avoid an event horizon, the redshift function must be finite everywhere, for $r>r_0$.

\subsection{Case-I}
With the shape function $b(r)=r_{0}+\rho_{0}r_{0}^{3}\ln\left(\frac{r_{0}}{r}\right)$ \cite{Rahaman16},  where $r_0$ is the size of the wormhole throat radius,  Eqs. \eqref{eqn.7} - \eqref{eqn.9} reduce to,
\begin{eqnarray}\label{eqn.10}
\rho&=&\left(-\frac{2\rho_{0}r_{0}^{3}}{\Upsilon {r^{3}}}\right)^{\frac{\alpha}{2}}\left(-\frac{\rho_{0}r_{0}^{3}}{r^{3}}\right)\\
p_{r}&=&\left(\frac{-2\rho_{0}r_{0}^{3}}{\Upsilon {r^{3}}}\right)^{\frac{\alpha}{2}}\left[\left(r_{0}+\rho_{0}r_{0}^{3}\ln\left(\frac{r_{0}}{r}\right) \right)  \left(\frac{9\alpha^{2}}{4r^{3}}+\frac{9\alpha}{4r^{3}}-\frac{1}{r^{3}}\right)+\frac{3\alpha \rho_{0}r_{0}^{3}}{4r^{3}}-\frac{9\alpha^{2}}{4r^{2}}-\frac{3\alpha}{2r^{2}}\right],\\  \label{eqn.11}
p_{t}&=&\left(\frac{-2\rho_{0}r_{0}^{3}}{\Upsilon {r^{3}}}\right)^{\frac{\alpha}{2}}\left[\left( r_{0}+\rho_{0}r_{0}^{3}\ln\left(\frac{r_{0}}{r}\right)\right) \left(\frac{1}{2r^{3}}-\frac{3\alpha}{2r^{3}}\right)+\frac{3\alpha}{2r^{2}}+\frac{\rho_{0}r_{0}^{3}}{2r^{3}}\right]. \label{eqn.12}
\end{eqnarray}
    
Now, the  energy conditions can be obtained as,
\begin{equation}
\rho+p_{r}=\left(\frac{-2\rho_{0}r_{0}^{3}}{\Upsilon {r^{3}}}\right)^{\frac{\alpha}{2}}\left[\left( r_{0}+\rho_{0}r_{0}^{3}\ln\left(\frac{r_{0}}{r}\right)\right) \left(\frac{9\alpha^{2}}{4r^{3}}+\frac{9\alpha}{4r^{3}}-\frac{1}{r^{3}}\right)+\frac{3\alpha \rho_{0}r_{0}^{3}}{4r^{3}}-\frac{9\alpha^{2}}{4r^{2}}-\frac{3\alpha}{2r^{2}}-\frac{\rho_{0}r_{0}^{3}}{r^{3}}\right],  \label{eqn.13}
\end{equation} 

\begin{equation}
\rho+p_{t}=\left(\frac{-2\rho_{0}r_{0}^{3}}{\Upsilon {r^{3}}}\right)^{\frac{\alpha}{2}}\left[\left( r_{0}+\rho_{0}r_{0}^{3}\ln\left(\frac{r_{0}}{r}\right)\right) \left(\frac{1}{2r^{3}}-\frac{3\alpha}{2r^{3}}\right)+\frac{3\alpha}{2r^{2}}-\frac{\rho_{0}r_{0}^{3}}{2r^{3}}\right], \label{eqn.14}
\end{equation}

\begin{equation}
p_{t}-p_{r}=\left(\frac{-2\rho_{0}r_{0}^{3}}{\Upsilon {r^{3}}}\right)^{\frac{\alpha}{2}}\left[\left( r_{0}+\rho_{0}r_{0}^{3}\ln\left(\frac{r_{0}}{r}\right)\right) \left(\frac{3}{2r^{3}}-\frac{15\alpha}{4r^{3}}-\frac{9\alpha^{2}}{4r^{3}}\right)+\frac{3\alpha}{r^{2}}+\frac{\rho_{0}r_{0}^{3}}{2r^{3}}-\frac{3\alpha\rho_{0}r_{0}^{3}}{4r^{3}}+\frac{9\alpha^{2}}{4r^{2}}\right], \label{eqn.15}
\end{equation}

\begin{equation}
\frac{p_r}{\rho}=\left[\frac{\left( r_{0}+\rho_{0}r_{0}^{3}\ln\left(\frac{r_{0}}{r}\right)\right) }{\rho_{0}r_{0}^{3}}\left(1-\frac{9\alpha^{2}}{4}-\frac{9\alpha}{4}\right)-\frac{3\alpha}{4}+\frac{9\alpha^{2}r}{4\rho_{0}r_{0}^{3}}+\frac{3r\alpha}{2\rho_{0}r_{0}^{3}}\right], \label{eqn.16}
\end{equation}

\begin{equation}
\rho+p_{r}+2p_{t}=\left(\frac{-2\rho_{0}r_{0}^{3}}{\Upsilon {r^{3}}}\right)^{\frac{\alpha}{2}}\left[\left(r_{0}+\rho_{0}r_{0}^{3}\ln\left(\frac{r_{0}}{r}\right)\right) \left(\frac{9\alpha^{2}}{4r^{3}}-\frac{3\alpha}{4r^{3}}\right)+\frac{3\alpha\rho_{0}r_{0}^{3}}{4r^{3}}-\frac{9\alpha^{2}}{4r^{2}}+\frac{3\alpha}{2r^{2}}\right]. \label{eqn.17}
\end{equation}

\begin{figure}[H]
\minipage{0.50\textwidth}
\centering
\includegraphics[width=\textwidth]{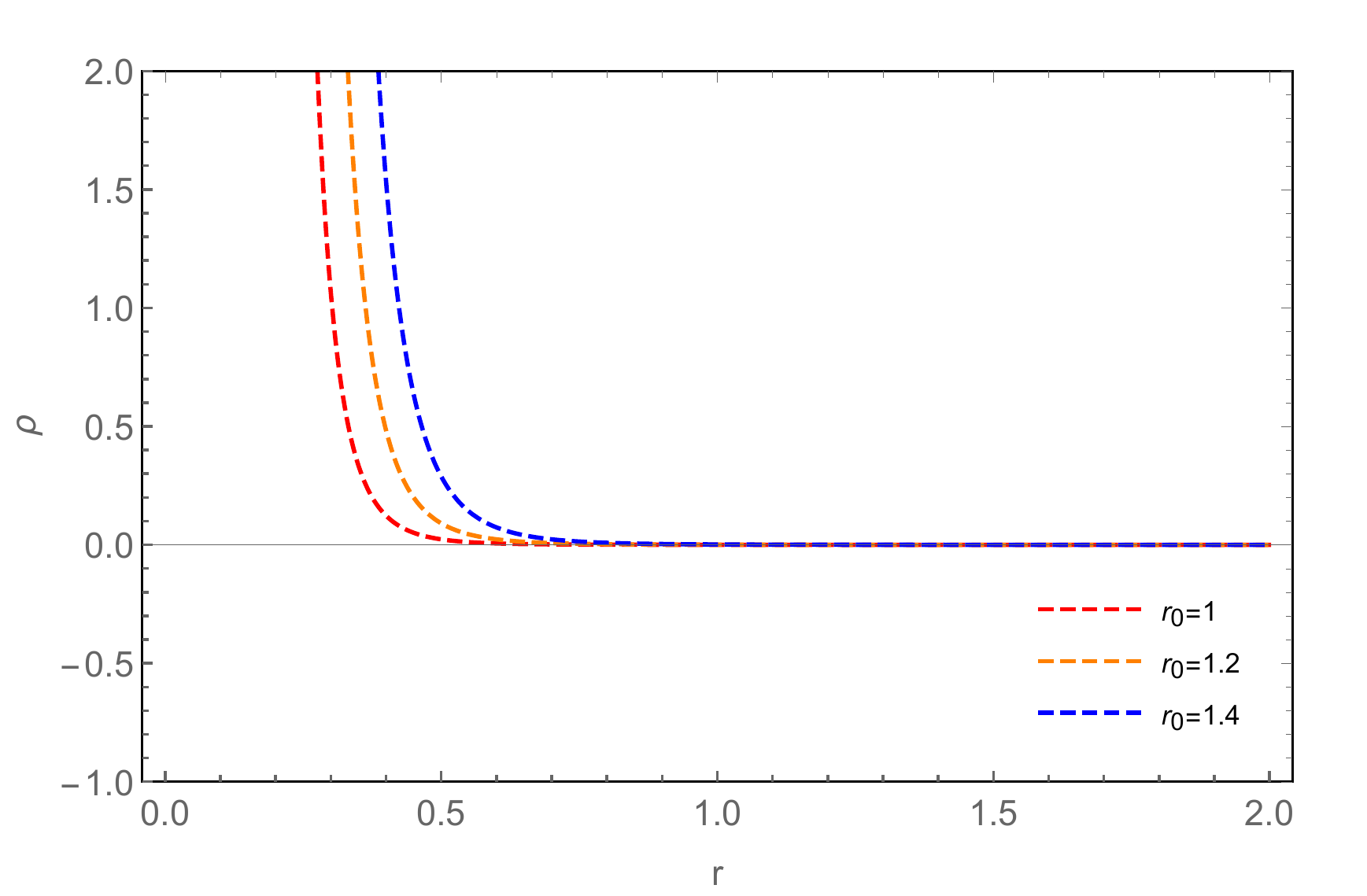}
\caption{$\rho$ vs $r$ (Case-I)}
\endminipage\hfill
\minipage{0.50\textwidth}
\includegraphics[width=\textwidth]{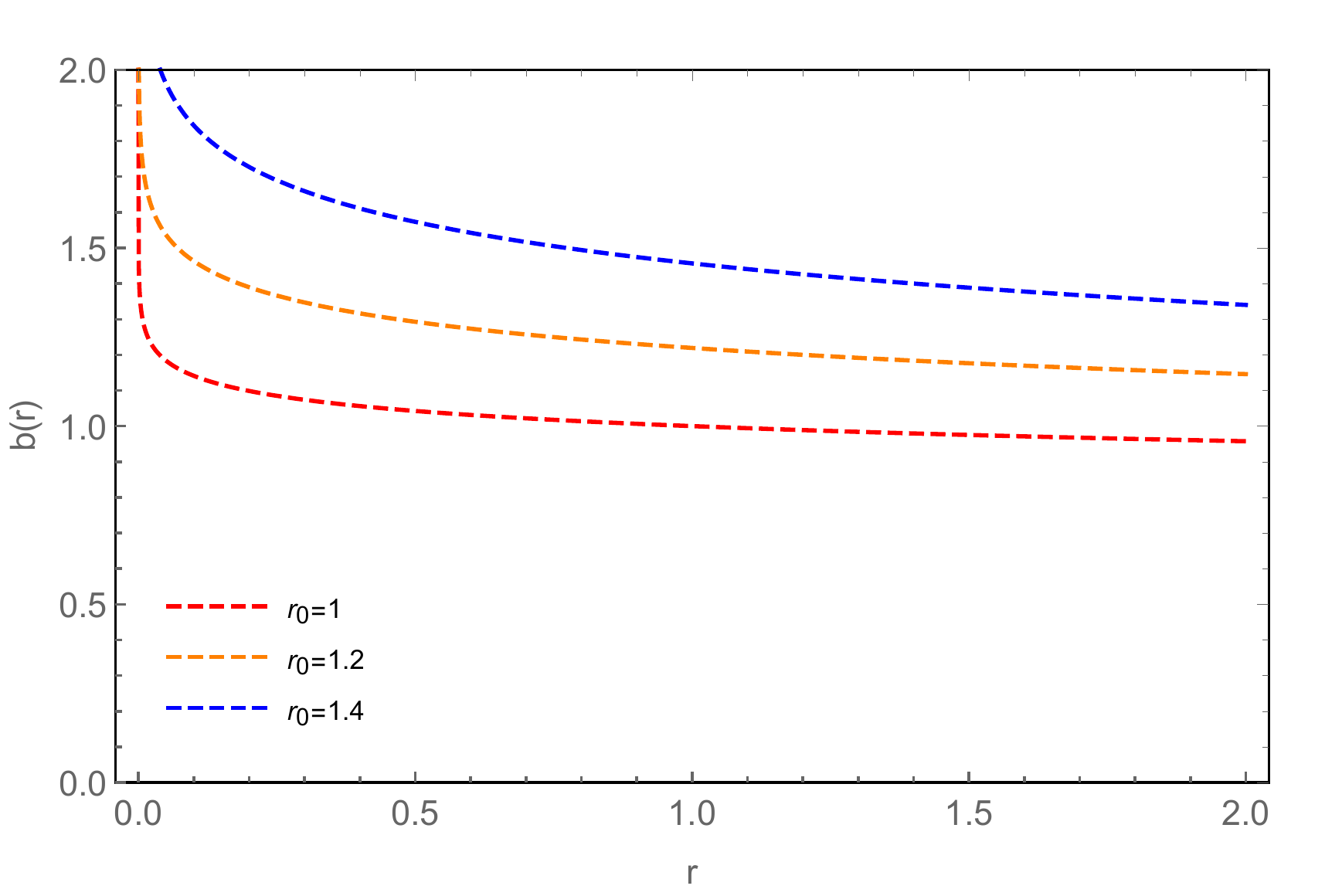} 
\caption{$b(r)$ vs $r$ (Case-I)}
\endminipage
\end{figure}

\begin{figure}[H]
\minipage{0.50\textwidth}
\centering
\includegraphics[width=\textwidth]{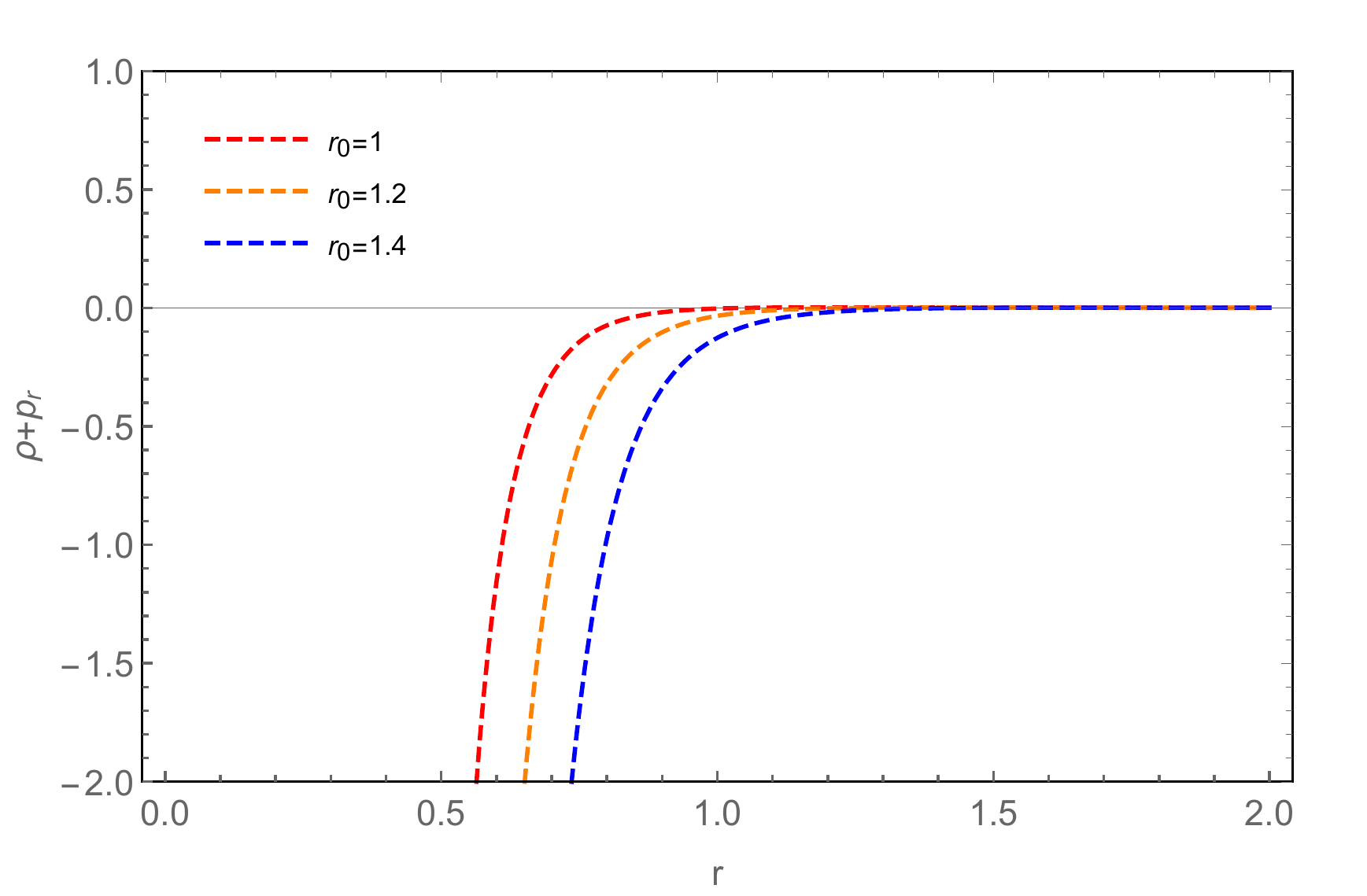}
\caption{$\rho+p_r$ vs $r$ (Case I)}
\endminipage\hfill
\minipage{0.50\textwidth}
\includegraphics[width=\textwidth]{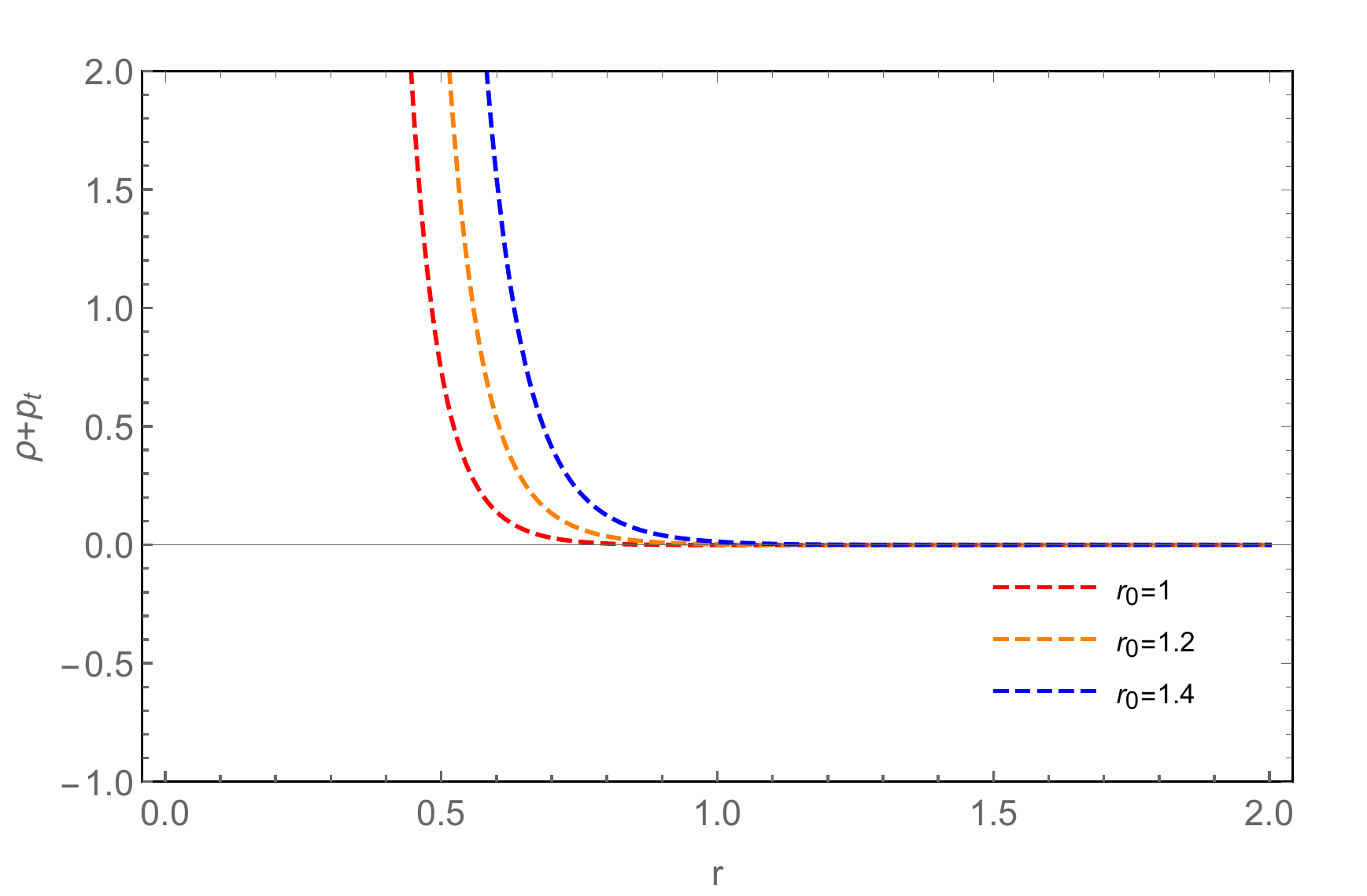} 
\caption{$\rho+p_t$ vs $r$ (Case I)}
\endminipage
\end{figure}

\begin{figure}[H]
\minipage{0.50\textwidth}
\centering
\includegraphics[width=\textwidth]{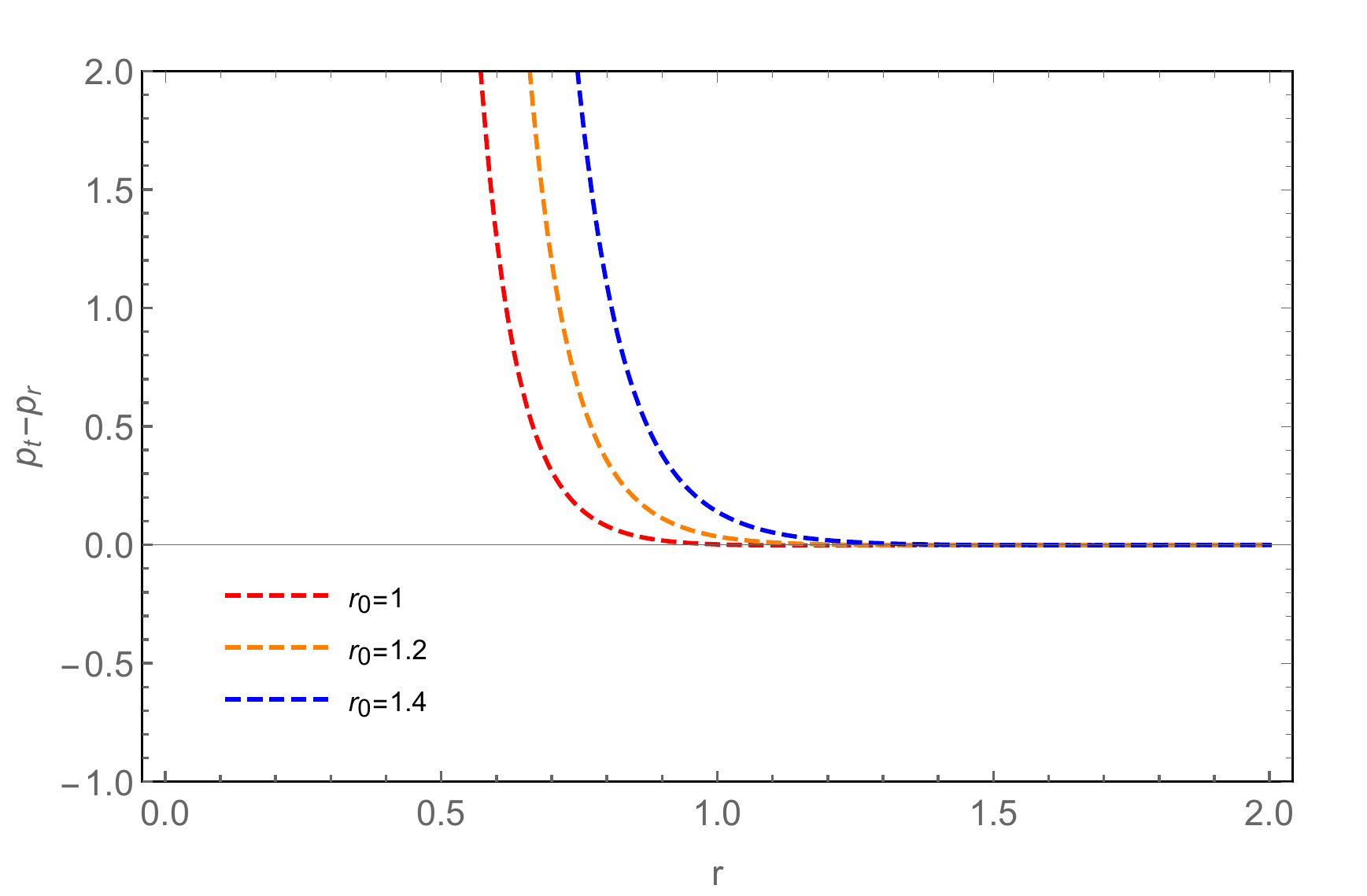}
\caption{$p_t-p_r$ vs $r$ (Case I)}
\endminipage\hfill
\minipage{0.50\textwidth}
\includegraphics[width=\textwidth]{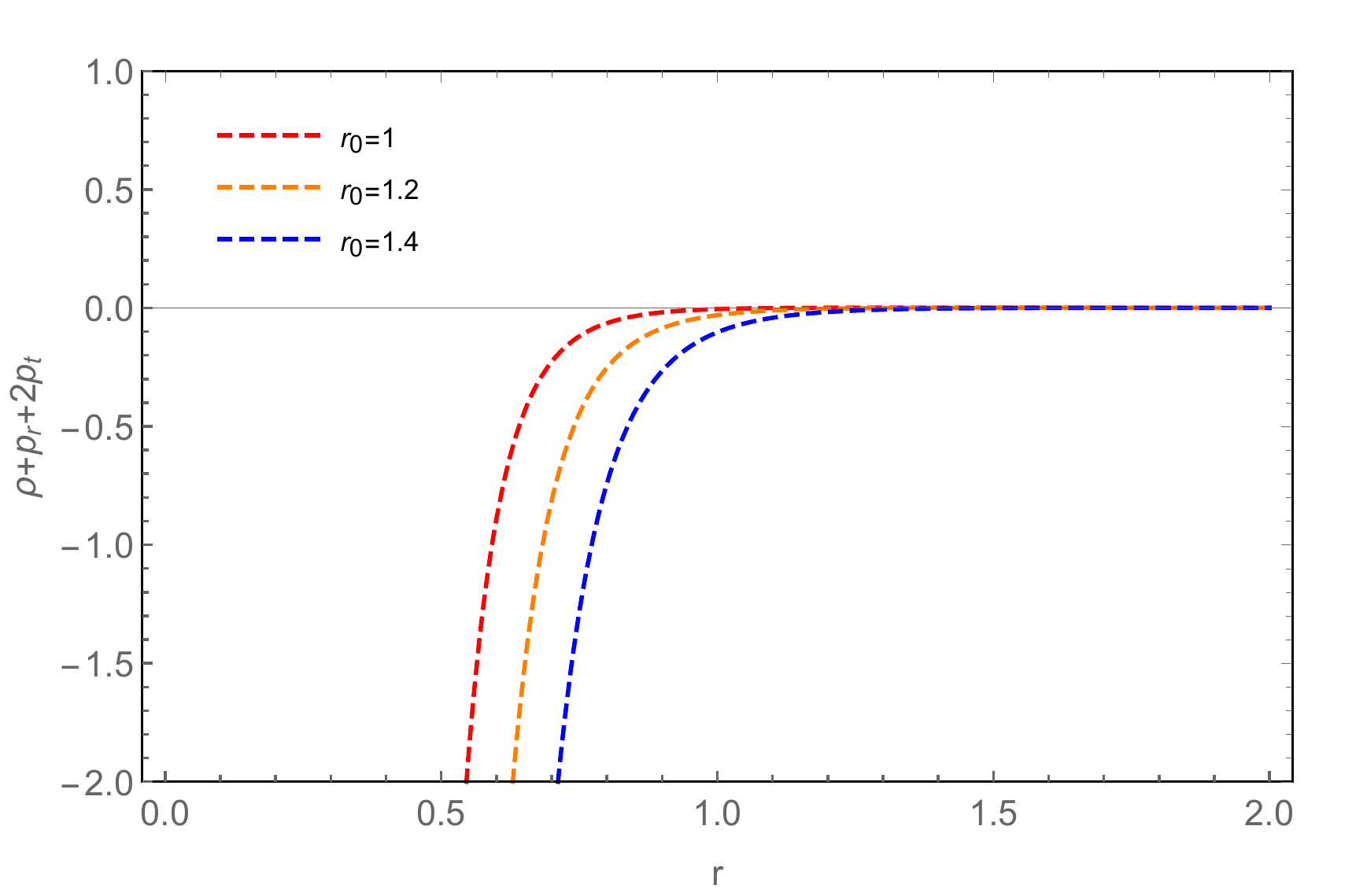} 
\caption{$\rho+p_r+2p_t$ vs $r$ (Case I)}
\endminipage
\end{figure}

\subsection{Case-II}
In this case, we consider the shape function as $b(r)=r_{0}+\gamma r_{0}\left(1-\dfrac{r_{0}}{r}\right)$ \cite{Rahaman16}. Now the energy density ($\rho$), radial ($p_r$) and tangential ($p_t$) pressures can be derived as,
\begin{eqnarray}\label{eqn.18}
\rho&=&\left(\frac{2\gamma r_{0}^{2}}{\Upsilon r^{4}}\right)^{\frac{\alpha}{2}}\left(\frac{\gamma r_{0}^{2}}{r^{4}}\right)\\
p_{r}&=&\left(\frac{2\gamma r_{0}^{2}}{\Upsilon r^{4}}\right)^{\frac{\alpha}{2}}\left[\left(r_{0}+\gamma r_{0}\left(1-\dfrac{r_{0}}{r}\right)\right)\left(-\frac{1}{r^{3}}+\frac{3\alpha}{r^{3}}+\frac{{4\alpha^{2}}}{r^{3}}\right)-\frac{\alpha \gamma r_{0}^{2}}{r^{4}}-\frac{{2\alpha}(2\alpha+1)}{r^{2}}\right],\\  \label{eqn.19}
p_{t}&=&\left(\frac{2\gamma r_{0}^{2}}{\Upsilon r^{4}}\right)^{\frac{\alpha}{2}}\left[\left(r_{0}+\gamma r_{0}\left(1-\dfrac{r_{0}}{r}\right)\right)\left(\frac{1}{2r^{3}}-\frac{2\alpha}{r^{3}}\right)+\frac{2\alpha}{r^{2}}-\frac{\gamma r_{0}^{2}}{2r^{4}}\right]. \label{eqn.20}
\end{eqnarray}

Hence, the energy conditions can be obtained as,
\begin{equation}
\rho+p_{r}=\left(\frac{2\gamma r_{0}^{2}}{\Upsilon r^{4}}\right)^{\frac{\alpha}{2}}\left[\left(r_{0}+\gamma r_{0}\left(1-\dfrac{r_{0}}{r}\right)\right) \left(-\frac{1}{r^{3}}+\frac{3\alpha}{r^{3}}+\frac{{4\alpha^{2}}}{r^{3}}\right)-\frac{\alpha \gamma r_{0}^{2}}{r^{4}}+\frac{\gamma r_{0}^{2}}{r^{4}}-\frac{{2\alpha}(2\alpha+1)}{r^{2}}\right],\label{eqn.21}
\end{equation}

\begin{equation}
\rho+p_{t}=\left(\frac{2\gamma r_{0}^{2}}{\Upsilon r^{4}}\right)^{\frac{\alpha}{2}}\left[\left(r_{0}+\gamma r_{0}\left(1-\dfrac{r_{0}}{r}\right)\right) \left(\frac{1}{2r^{3}}-\frac{2\alpha}{r^{3}}\right)+\frac{2\alpha}{r^{2}}+\frac{\gamma r_{0}^{2}}{2r^{4}}\right],
\end{equation}\label{eqn.22}

\begin{equation}
p_{t}-p_{r}=\left(\frac{2\gamma r_{0}^{2}}{\Upsilon r^{4}}\right)^{\frac{\alpha}{2}}\left[\left(r_{0}+\gamma r_{0}\left(1-\dfrac{r_{0}}{r}\right)\right)  \left(\frac{3}{2r^{3}}-\frac{5\alpha}{r^{3}}-\frac{{4\alpha^{2}}}{r^{3}}\right)+\frac{\alpha \gamma r_{0}^{2}}{r^{4}}-\frac{\gamma r_{0}^{2}}{2r^{4}}+\frac{{2\alpha}(2\alpha+2)}{r^{2}}\right], \label{eqn.23}
\end{equation}

\begin{equation}
\frac{p_r}{\rho}=\left[\frac{\left(r_{0}+\gamma r_{0}\left(1-\dfrac{r_{0}}{r}\right)\right)  r}{\gamma r_{0}^{2}}\left(3\alpha+4\alpha^{2}-1\right)-\alpha-\frac{{2\alpha}(2\alpha+1)r^{2}}{\gamma r_{0}^{2}}\right], \label{eqn.24}
\end{equation}

\begin{equation}
\rho+p_{r}+2p_{t}=\left(\frac{2\gamma r_{0}^{2}}{\Upsilon r^{4}}\right)^{\frac{\alpha}{2}}\left[\left(r_{0}+\gamma r_{0}\left(1-\dfrac{r_{0}}{r}\right)\right)\left(\frac{4\alpha^{2}}{r^{3}}-\frac{\alpha}{r^{3}}\right)-\frac{\alpha \gamma r_{0}^{2}}{r^{4}}-\frac{4\alpha^{2}}{r^{2}}+\frac{2\alpha}{r^{2}}\right]. \label{eqn.25}
\end{equation}

\begin{figure}[H]
\minipage{0.50\textwidth}
\centering
\includegraphics[width=\textwidth]{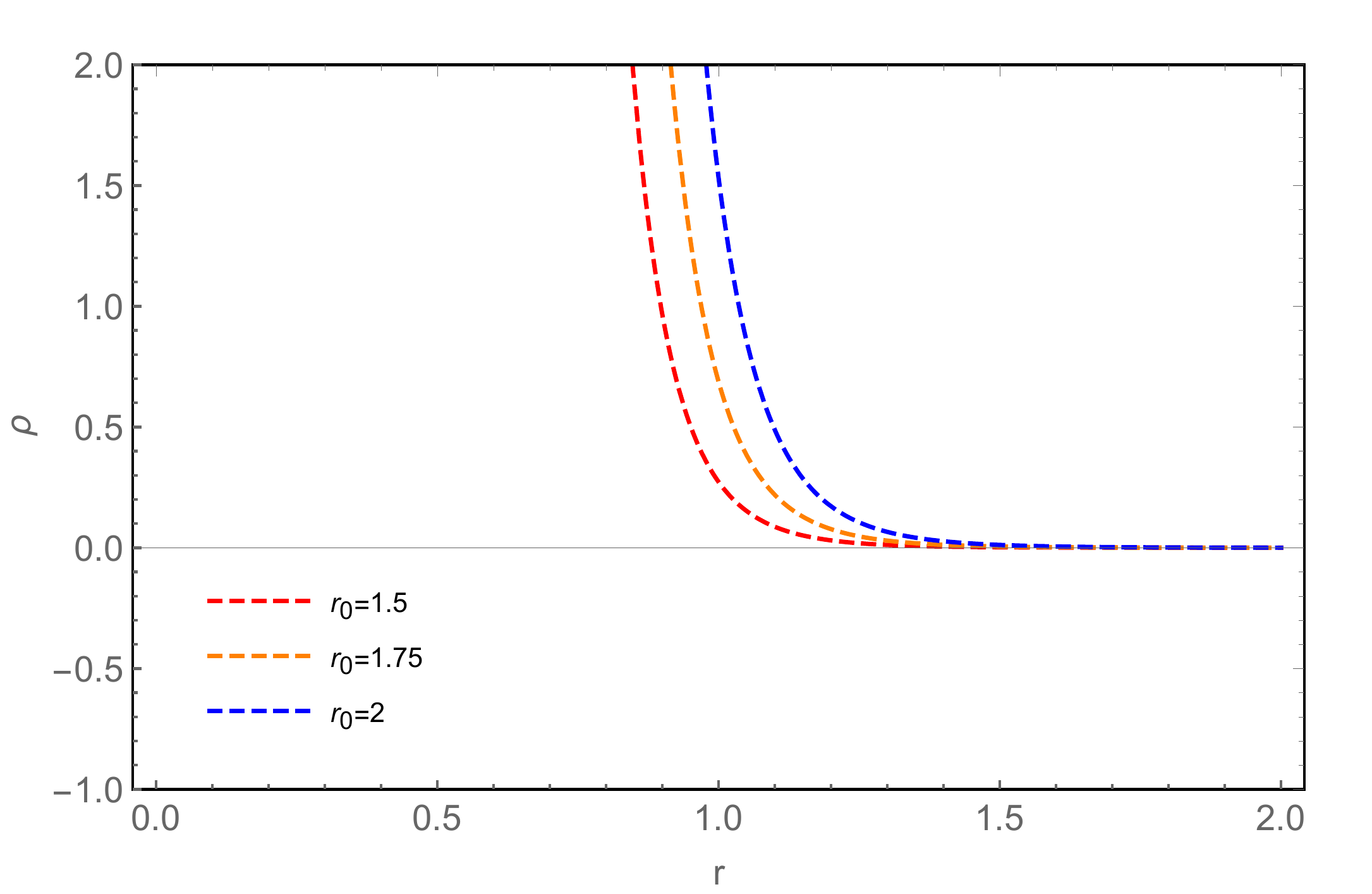}
\caption{$\rho$ vs $r$ (Case II)}
\endminipage\hfill
\minipage{0.50\textwidth}
\includegraphics[width=\textwidth]{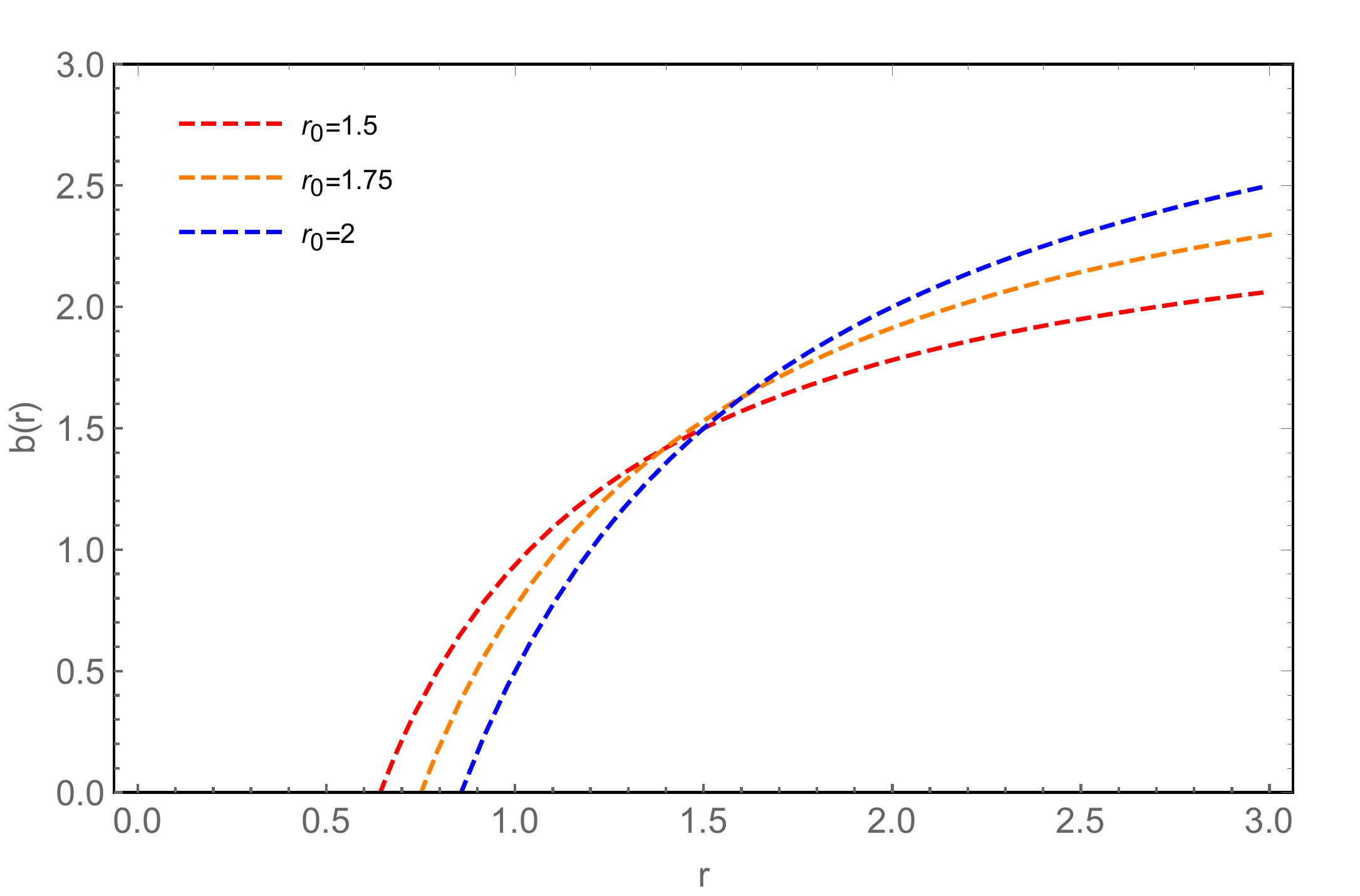} 
\caption{$b(r)$ vs $r$ (Case II)}
\endminipage
\end{figure}

\begin{figure}[H]
\minipage{0.50\textwidth}
\centering
\includegraphics[width=\textwidth]{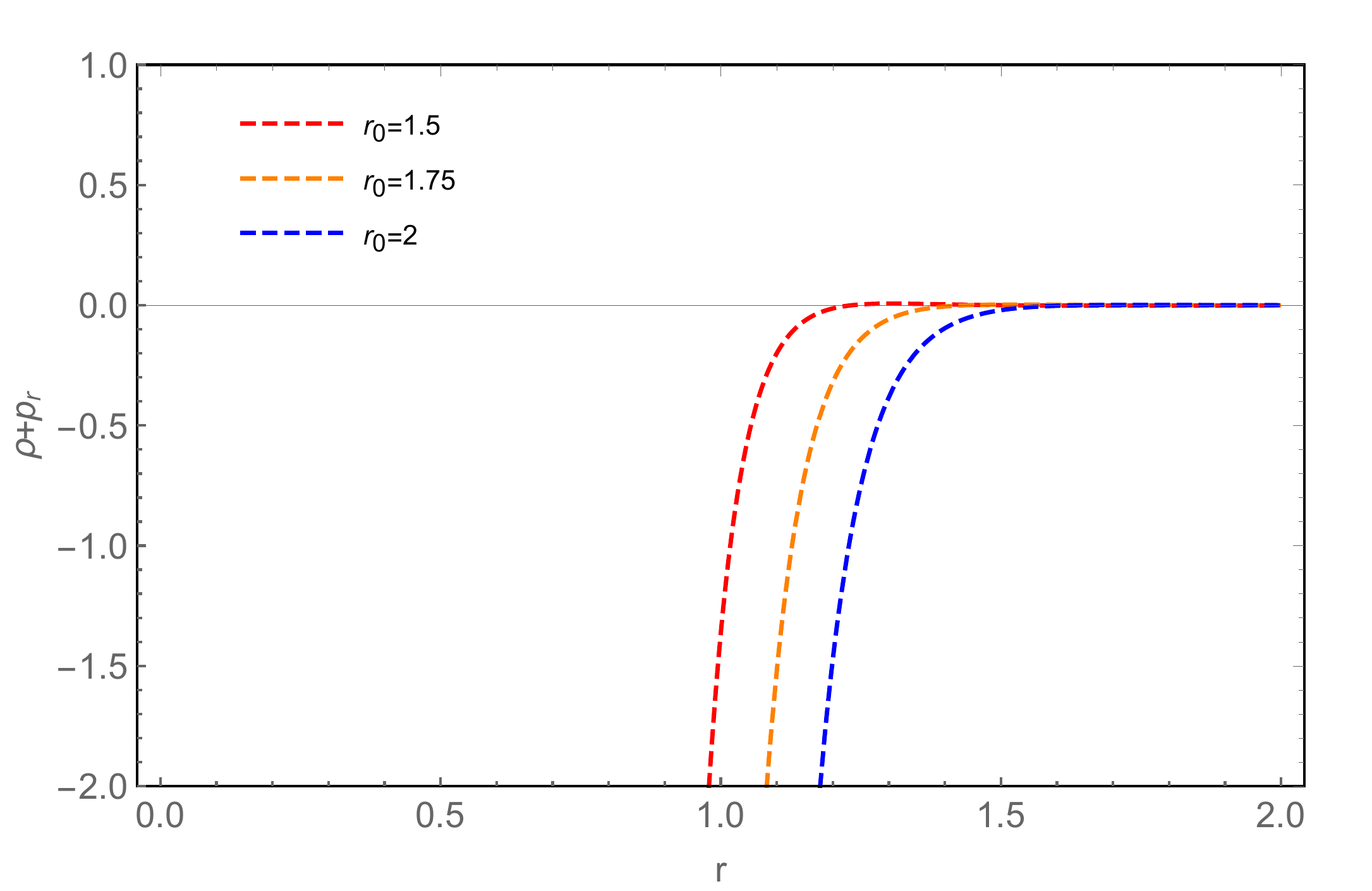}
\caption{$\rho+p_r$ vs $r$ (Case II)}
\endminipage\hfill
\minipage{0.50\textwidth}
\includegraphics[width=\textwidth]{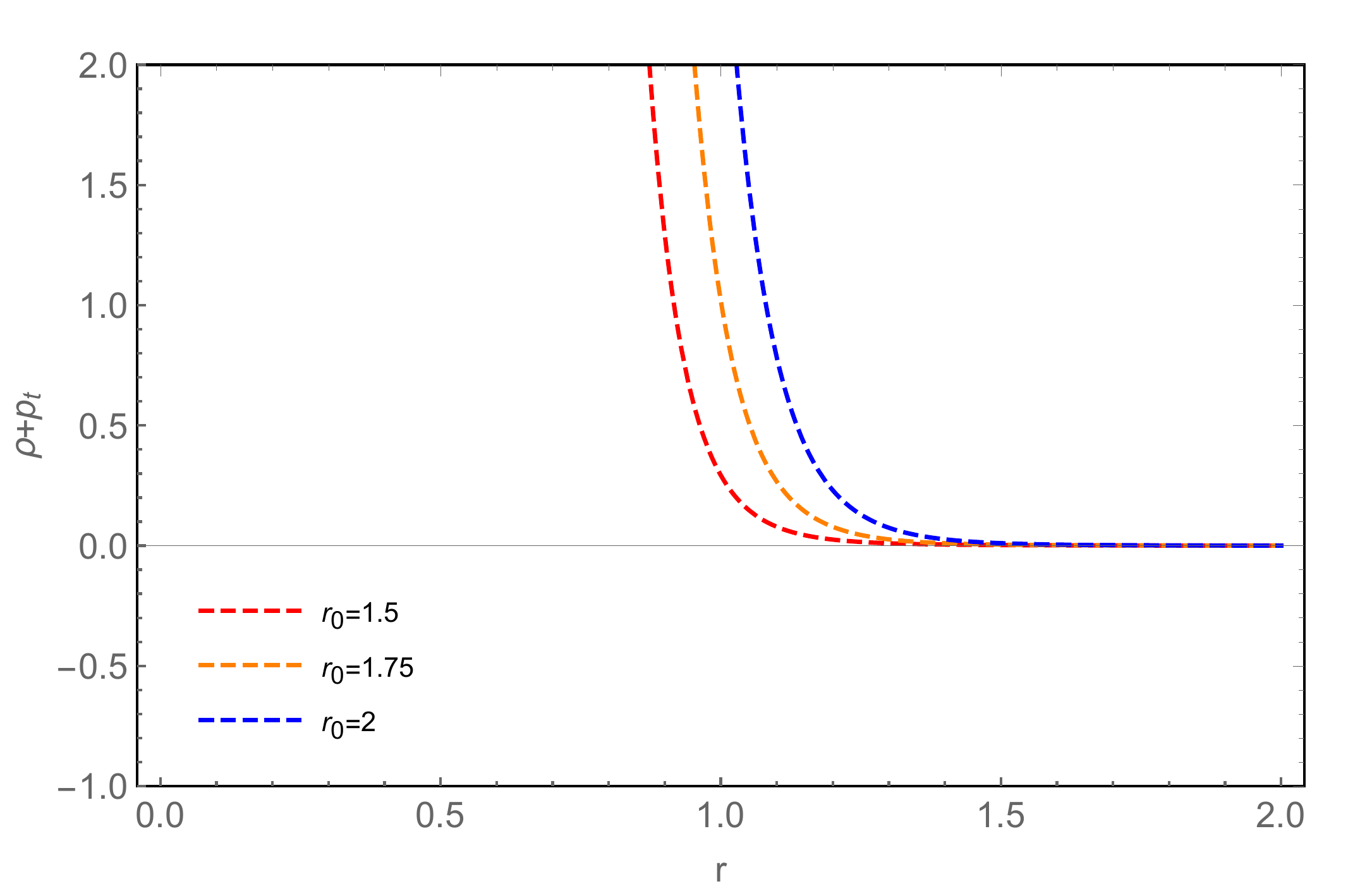} 
\caption{$\rho+p_t$ vs $r$ (Case II)}
\endminipage
\end{figure}

\begin{figure}[H]
\minipage{0.50\textwidth}
\centering
\includegraphics[width=\textwidth]{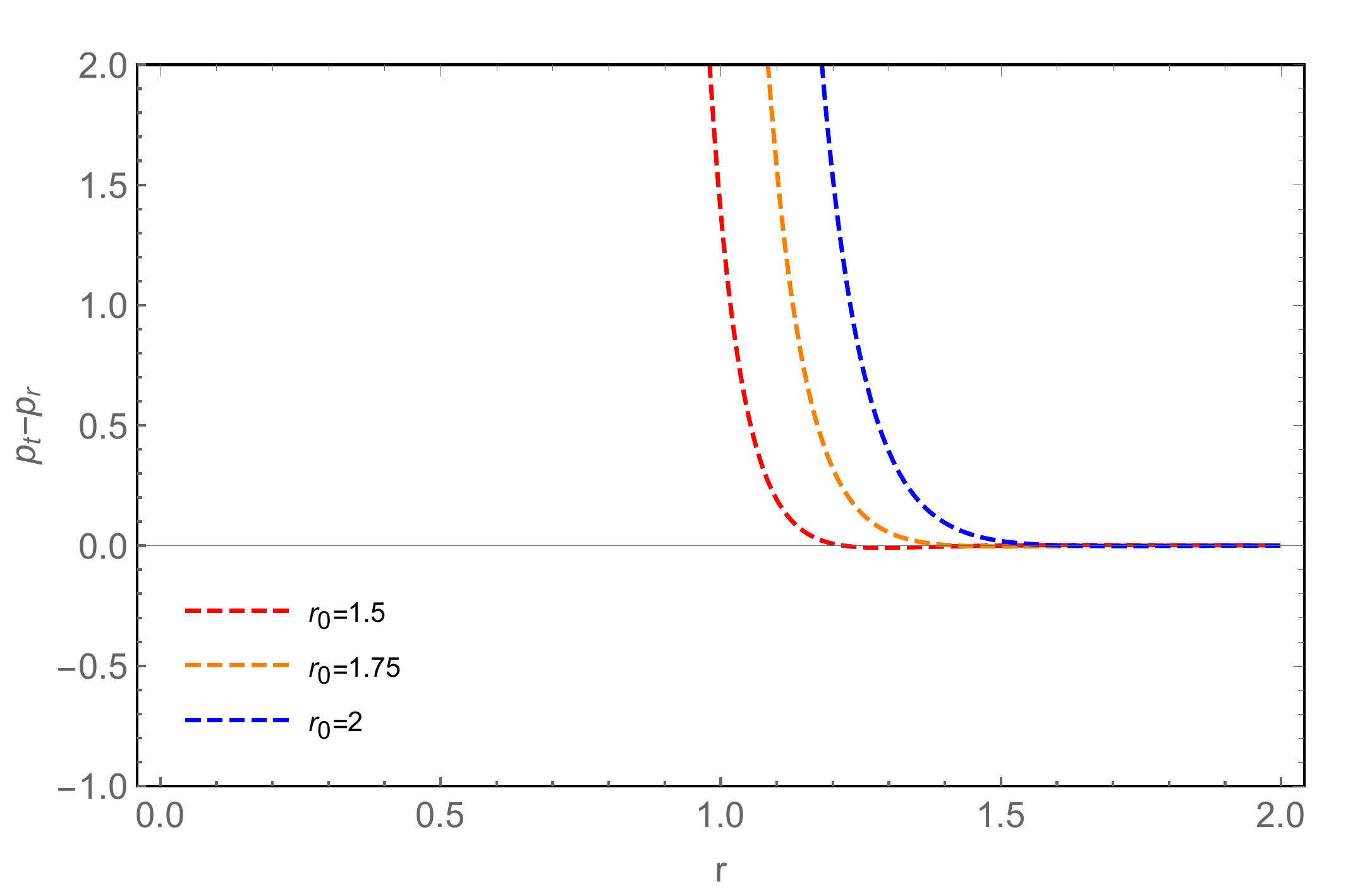}
\caption{$p_t-p_r$ vs $r$ (Case II)}
\endminipage\hfill
\minipage{0.50\textwidth}
\includegraphics[width=\textwidth]{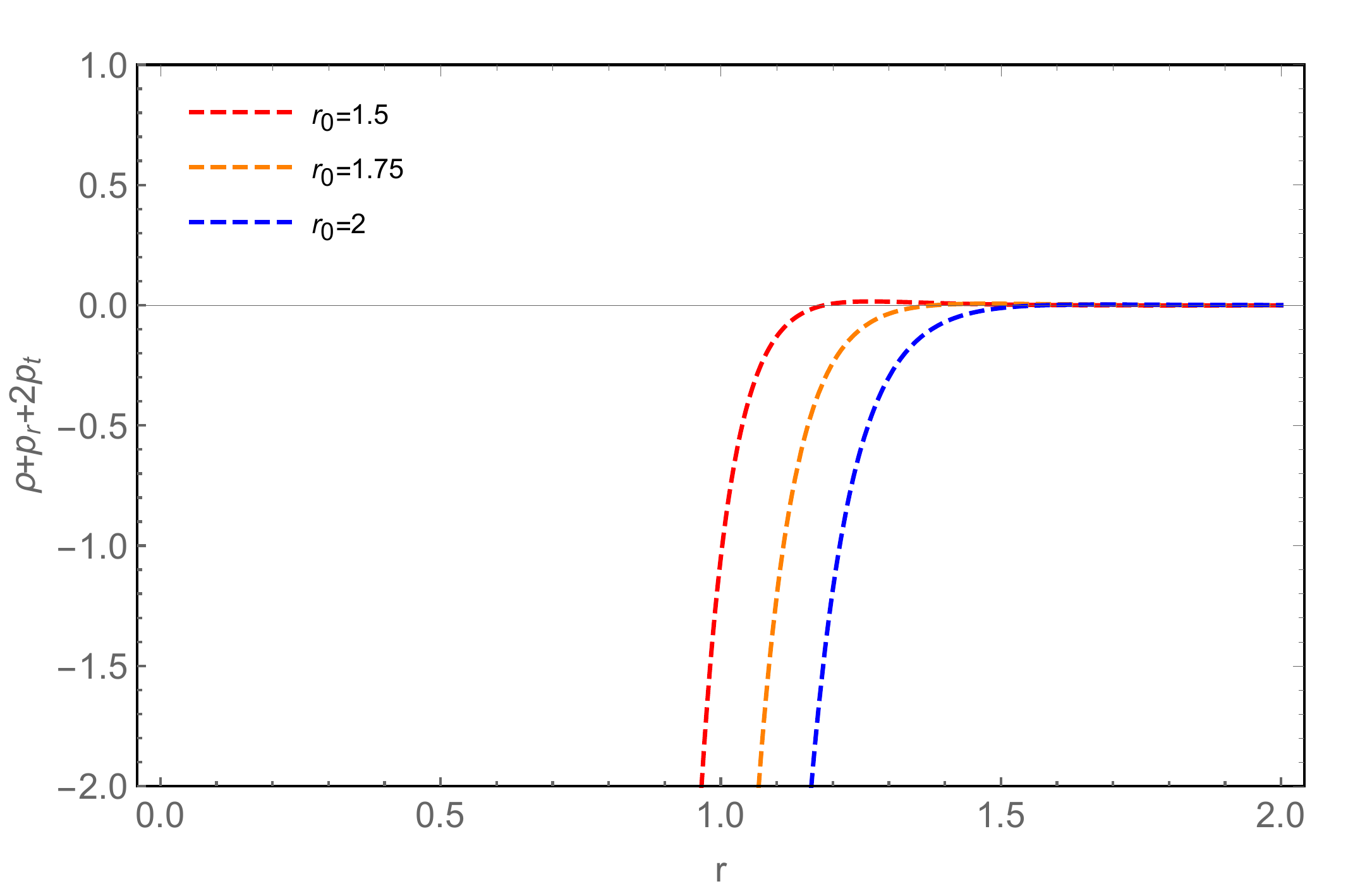} 
\caption{$\rho+p_r+2p_t$ vs $r$ (Case II)}
\endminipage
\end{figure}

\subsection{Case-III}
In this case, we consider the shape function as $b(r)=\alpha +\beta r$ \cite{Cataldo17}, and with this, the energy density $\rho$, radial component ($p_r$) and tangential component ($p_t$) of the pressure can be derived as,
\begin{eqnarray}\label{eqn.26}
\rho&=&\left(\frac{2\beta}{\Upsilon r^{2}}\right)^{\frac{\alpha}{2}}\dfrac{\beta}{r^{2}},\\
p_{r}&=&\left(\frac{2\beta}{\Upsilon r^{2}}\right)^{\frac{\alpha}{2}}\left[\left( \alpha^{2}+\dfrac{3\alpha}{2}-1\right)\dfrac{r_{0}(1-\beta)}{r^{3}}+\frac{(\beta-1)\left( \alpha^{2}+\alpha\right)}{r^{2}}-\frac{\beta}{r^{2}}\right],\\ \label{eqn.27}
p_{t}&=&\left(\frac{2\beta}{\Upsilon r^{2}}\right)^{\frac{\alpha}{2}}\left[\dfrac{r_{0}(1-\beta)}{r^{3}}\left(\dfrac{1}{2}-\alpha\right)-\dfrac{\alpha \beta}{r^{2}}+\dfrac{\alpha}{r^{2}} \right].\label{eqn.28}
\end{eqnarray} 

Now, the energy conditions can be obtained as, 
\begin{equation}
\rho+p_{r}=\left(\dfrac{2\beta}{\Upsilon r^{2}}\right)^{\frac{\alpha}{2}}\left[\left( \alpha^{2}+\dfrac{3\alpha}{2}-1\right)\dfrac{r_{0}(1-\beta)}{r^{3}}+\frac{(\beta-1)\left( \alpha^{2}+\alpha\right)}{r^{2}}\right],\label{eqn.29}
\end{equation}

\begin{equation}
\rho+p_{t}=\left(\dfrac{2\beta}{\Upsilon r^{2}}\right)^{\frac{\alpha}{2}}\left[\dfrac{r_{0}(1-\beta)}{r^{3}}\left(\dfrac{1}{2}-\alpha\right)-\dfrac{\alpha \beta}{r^{2}}+\dfrac{\alpha}{r^{2}}+\dfrac{\beta}{r^{2}} \right],\label{eqn.30}
\end{equation}

\begin{equation}
p_{t}-p_{r}=\left(\dfrac{2\beta}{\Upsilon r^{2}}\right)^{\frac{\alpha}{2}}\left[\left( \dfrac{3}{2}-\dfrac{5\alpha}{2}-\alpha^{2}\right)\dfrac{r_{0}(1-\beta)}{r^{3}}+\dfrac{(\alpha^{2}+2\alpha)(1-\beta)}{r^{2}}+\dfrac{\beta}{r^{2}}\right],\label{eqn.31}
\end{equation}

\begin{equation}
\frac{p_r}{\rho}=\dfrac{1}{\beta}\left[\left( \alpha^{2}+\dfrac{3\alpha}{2}-1\right)\dfrac{r_{0}(1-\beta)}{r}+(\beta-1)\left( \alpha^{2}+\alpha\right)-\beta\right], \label{eqn.32}
\end{equation}

\begin{equation}
\rho+p_{r}+2p_{t}=\left(\dfrac{2\beta}{\Upsilon r^{2}}\right)^{\frac{\alpha}{2}}\left[\left( \alpha^{2}-\dfrac{\alpha}{2}\right)\dfrac{r_{0}(1-\beta)}{r^{3}}+\dfrac{(\beta-1)(\alpha^{2}-\alpha )}{r^{2}}\right].\label{eqn.33}
\end{equation}

\begin{figure}[H]
\minipage{0.50\textwidth}
\centering
\includegraphics[width=\textwidth]{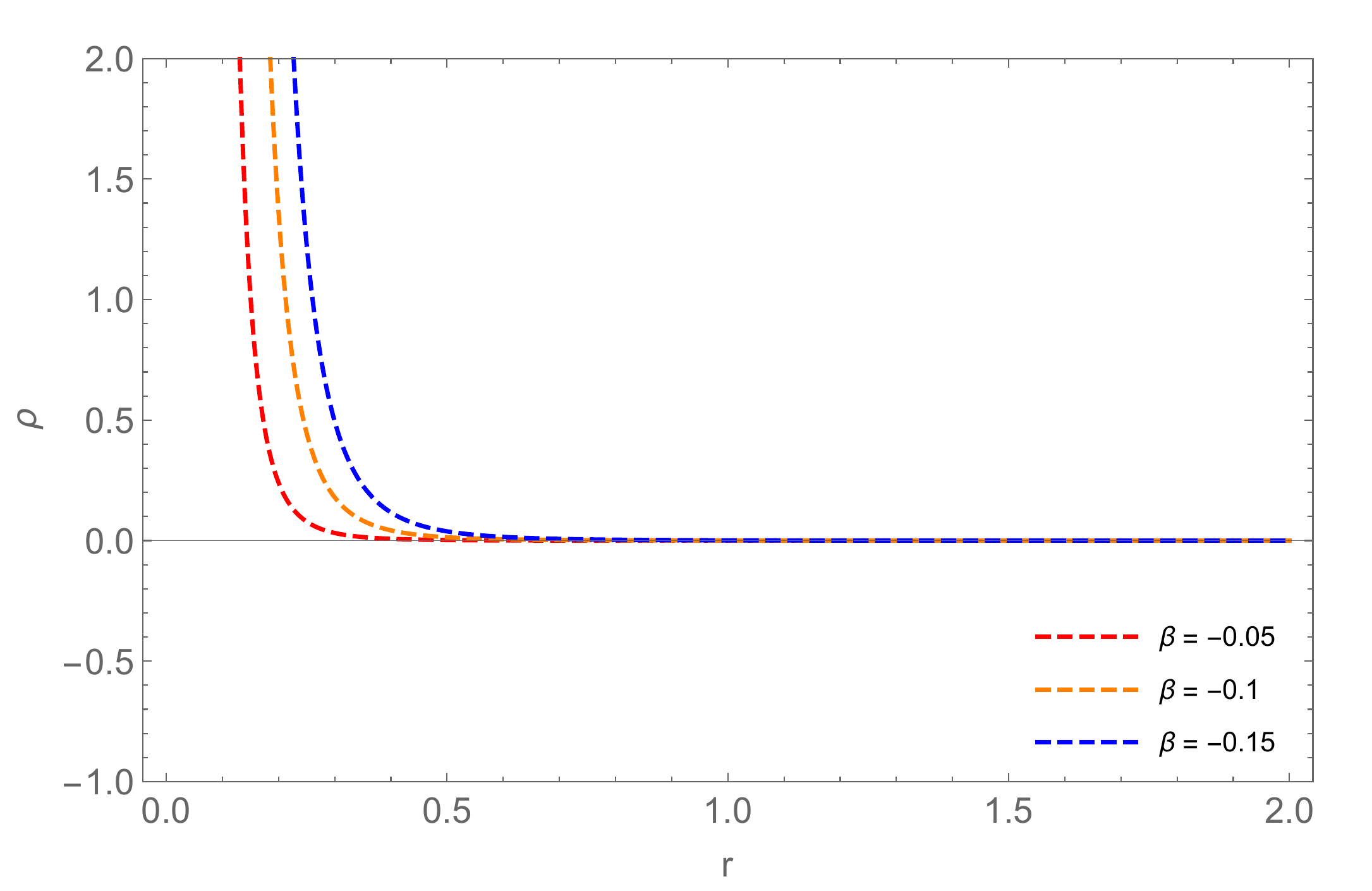}
\caption{$\rho$ vs $r$ (Case III)}
\endminipage\hfill
\minipage{0.50\textwidth}
\includegraphics[width=\textwidth]{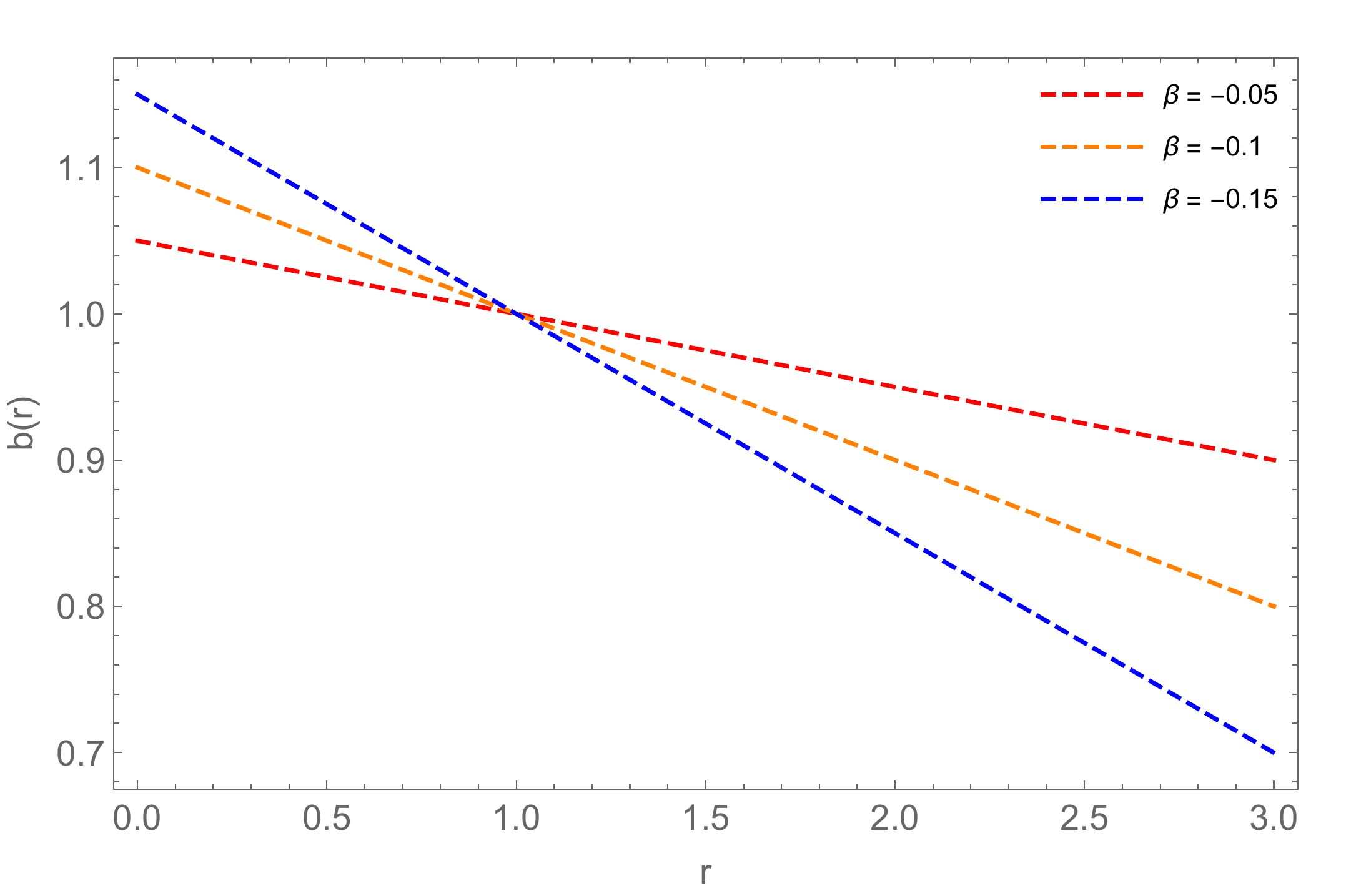} 
\caption{$b(r)$ vs $r$ (Case III)}
\endminipage
\end{figure}

\begin{figure}[H]
\minipage{0.50\textwidth}
\centering
\includegraphics[width=\textwidth]{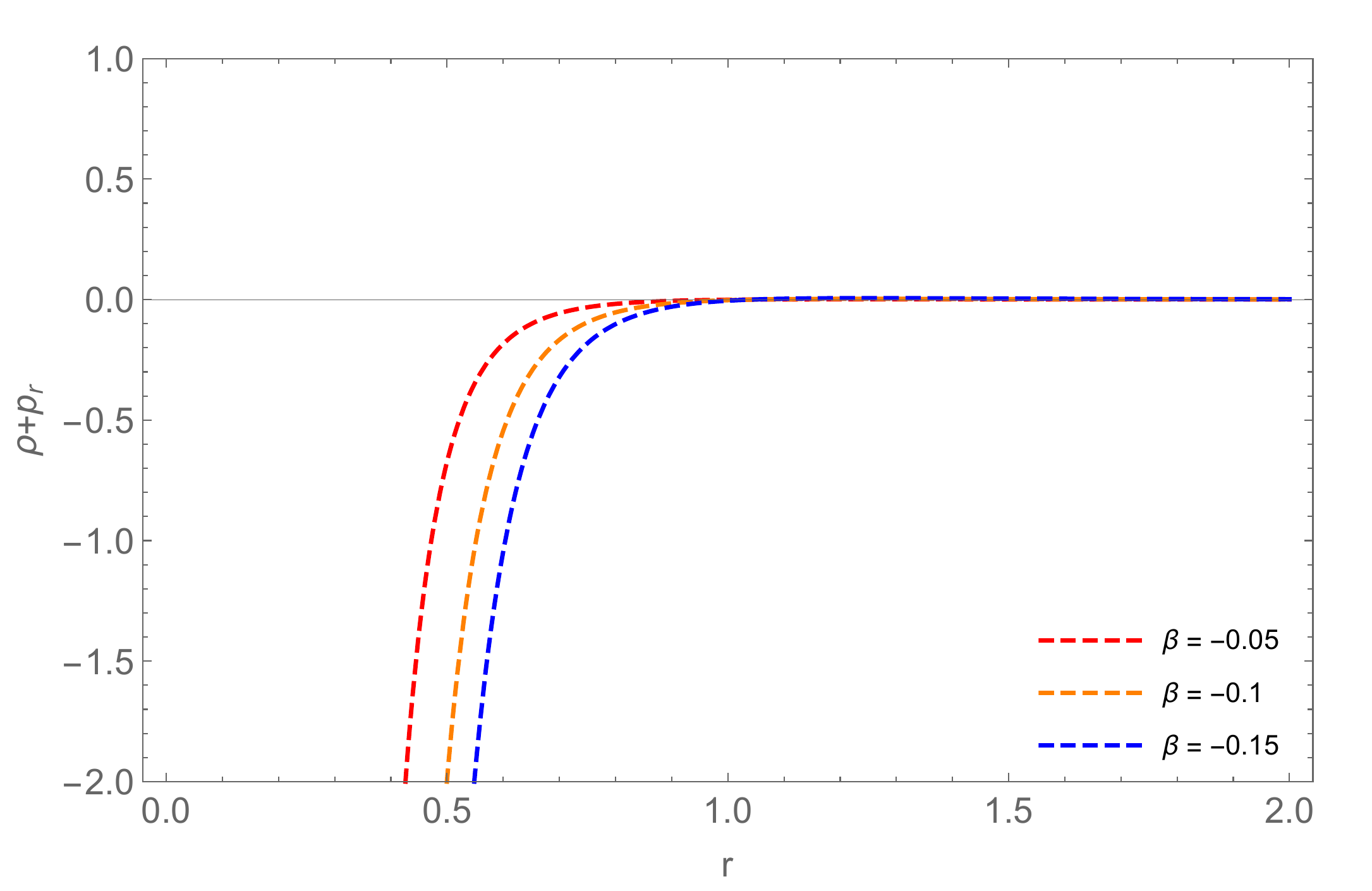}
\caption{$\rho+p_r$ vs $r$ (Case III)}
\endminipage\hfill
\minipage{0.50\textwidth}
\includegraphics[width=\textwidth]{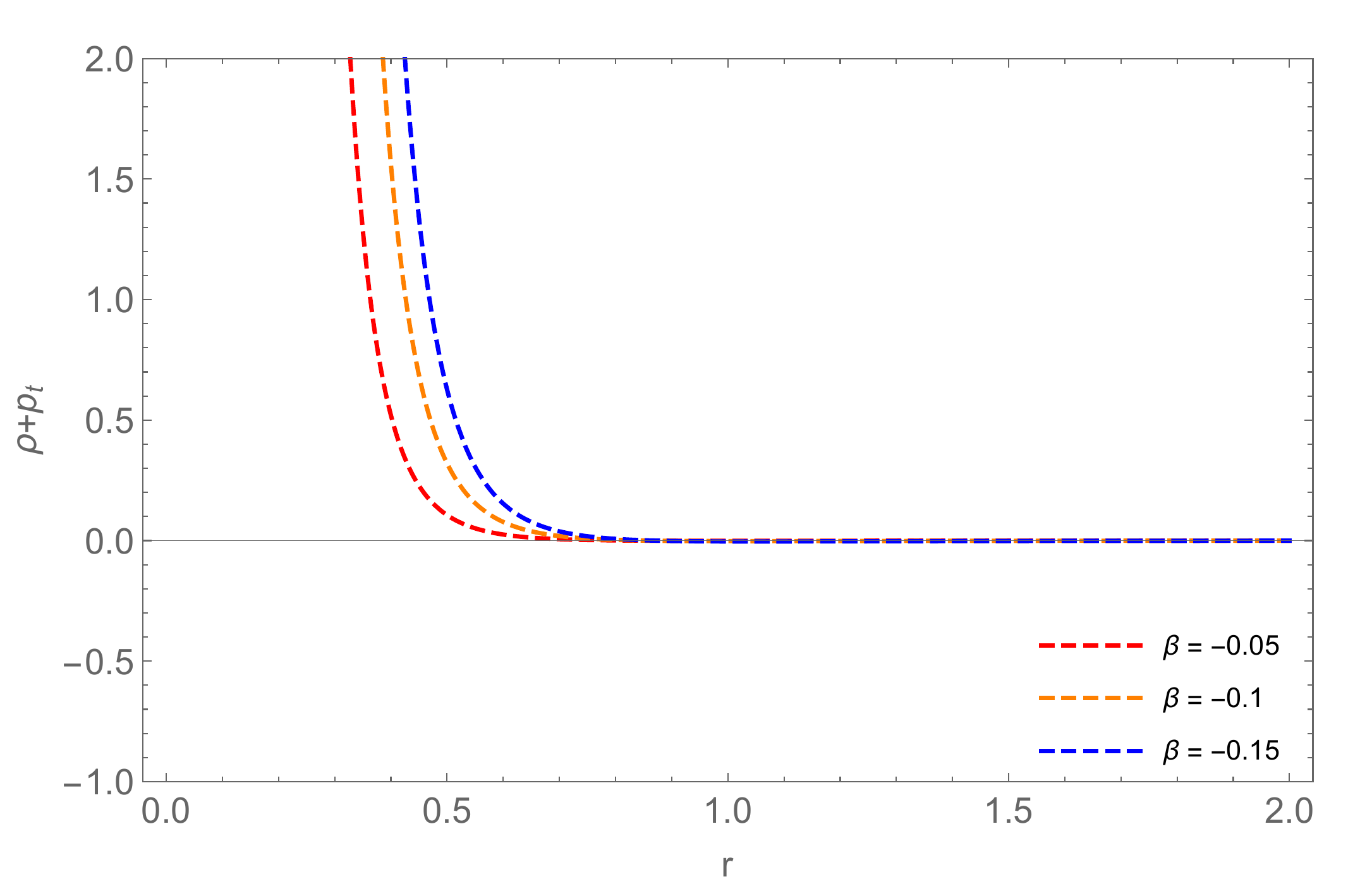} 
\caption{$\rho+p_t$ vs $r$ (Case III)}
\endminipage
\end{figure}

\begin{figure}[H]
\minipage{0.50\textwidth}
\centering
\includegraphics[width=\textwidth]{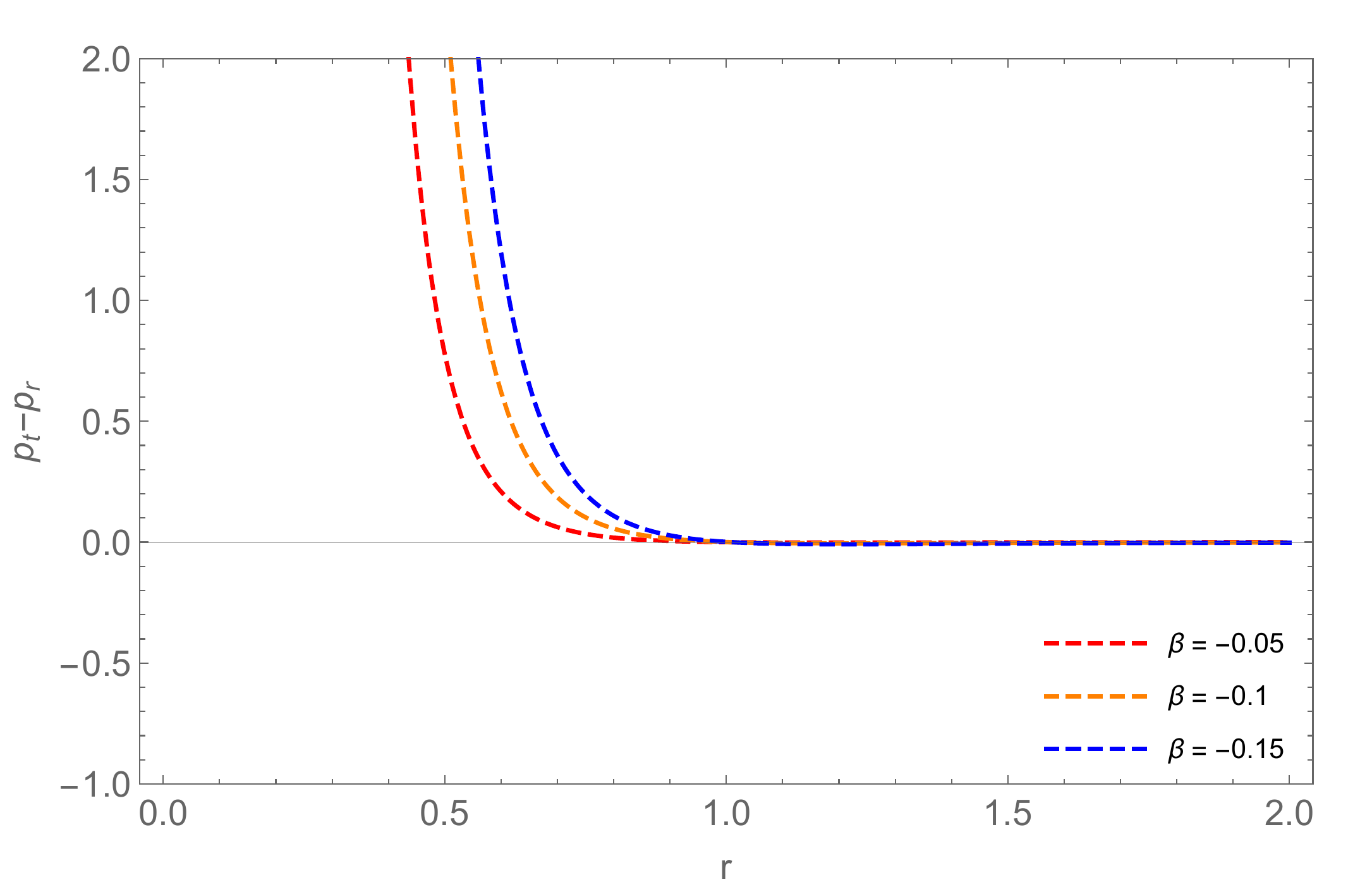}
\caption{$p_t-p_r$ vs $r$ (Case III)}
\endminipage\hfill
\minipage{0.50\textwidth}
\includegraphics[width=\textwidth]{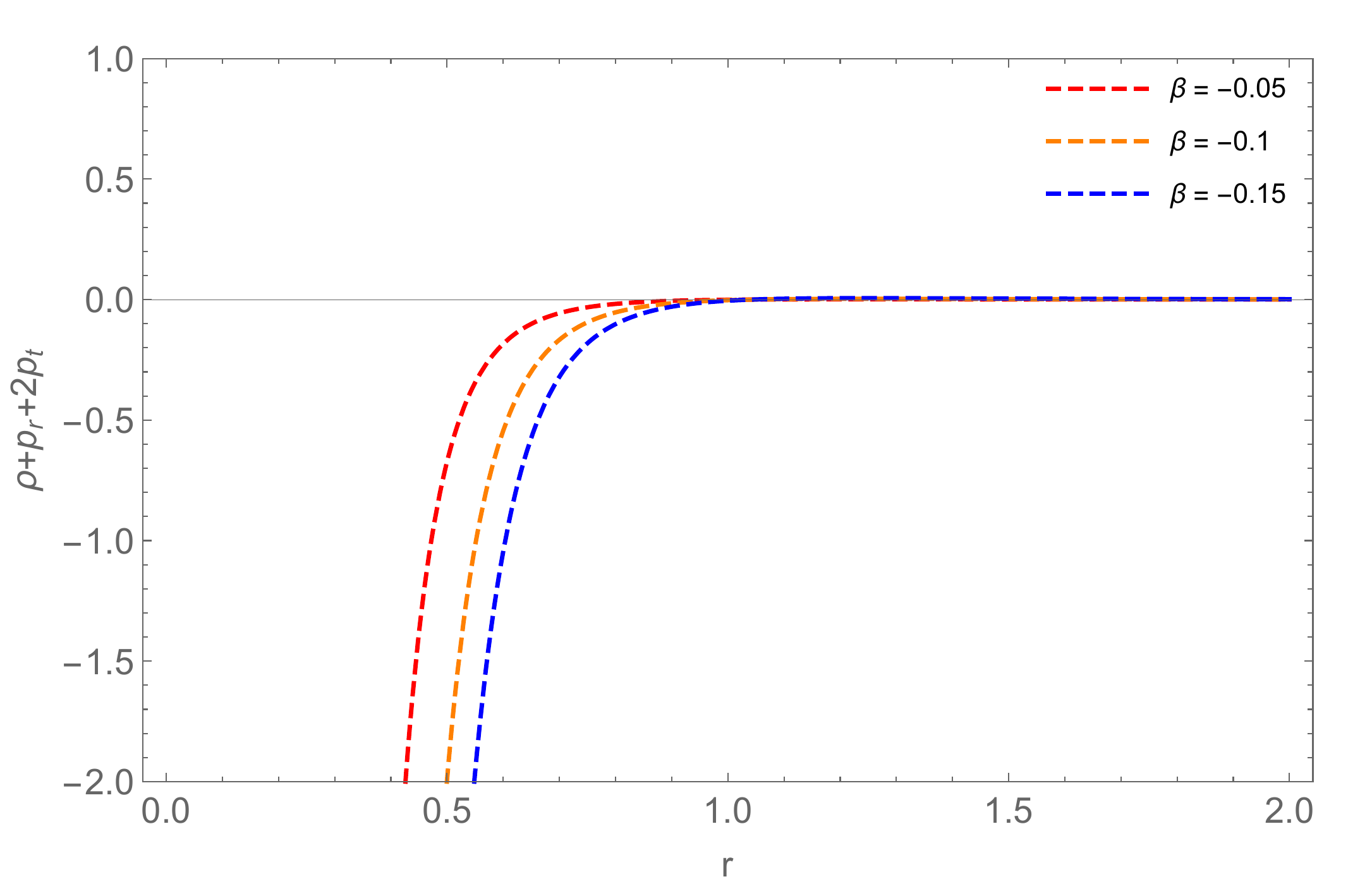} 
\caption{$\rho+p_r+2p_t$ vs $r$ (Case III)}
\endminipage
\end{figure}

\section{Discussions and Conclusion}
Our sole aim in the present investigation is to construct traversable wormhole models in $f(R)$ gravity. For this study we have specifically assumed the following three shape functions: (i) $b(r)=r_{0}+\rho_{0}r_{0}^{3}\ln\left(\frac{r_{0}}{r}\right)$, (ii) $b(r)=r_{0}+\gamma r_{0}\left(1-\dfrac{r_{0}}{r}\right)$, and (iii) $b(r)=\alpha +\beta r$. It is observed from the graphical plots that, by imposing some physical conditions on the parameters as well as constants involved here with in the shape functions, the behaviour of the physical parameters are interesting and viable. To explore features from the results as exhibited in the graphical plots we put here the following discussions:\\

(i) {\bf Plots for $\rho$ and $b(r)$ (Case I)}: The graphical representation of the shape function and energy conditions are presented with the representative value of $r_0=1,~1.2,~1.4$. The anisotropic parameter and model parameter having values respectively $k=0.95$ and $\alpha=3$ and the condition imposed on the shape function $b'(r_{0})\leq 1$ implies $\rho_{0} > 0$, hence at $\rho_0=0.06125$. The shape function decreases from higher values and with the increasing value of radial coordinate $r$, tending to one. The NEC$_r$ (which is on the radial pressure direction) for all the representative value showing the violation of null energy condition, whereas NEC$_t$ (which is on the tangential pressure direction) fails to violate the null energy condition. The difference between the tangential and radial pressure $p_t-p_r$ decreasing from a higher positive value and vanishes with the increase in the value of radial coordinate. The graphical behaviour of energy condition $\rho+p_r+2p_t$ shows the violation of energy condition. \\  

(ii) {\bf Plots for $\rho$ and $b(r)$ (Case II)}: The graphical representation of shape function and other energy conditions with representative value of $r_0$. The condition imposed on the shape function $b'(r_0)\leq 1$ implies $\gamma \leq 1$ and the condition $[b(r)-b'(r)r]/2b^{2}(r)\geq 0$ implies $\gamma \geq -1$ (i.e., $-1<\gamma \leq 1$). Thus at $\gamma=0.75$ with the value of anisotropic parameter and model parameter are respectively $k=0.95$ and $\alpha=4$. The shape function increases indefinitely with increasing value of radial coordinate. The NEC$_r$  violates the null energy condition whereas NEC$_t$ fails to violate null energy condition. This might be due to the different behaviour of radial pressure and tangential pressure are behaving differently, however both will vanish while the value increases. The difference between the pressure term decreases from a higher value and vanishes with the increase in the value of $r$. At the same time $\rho+p_r+2p_t$ violates the energy condition.\\

(iii){\bf Plots for $\rho$ and $b(r)$ (Case III)}: The shape function is a solution of a wormhole equation should satisfies $b(r_0)=r_0$, implies $b(r)=r_{0}(1-\beta)+\beta r$. The graphical representation of shape function and null energy condition with the representative value of $\beta$, condition imposed on the shape function $b(r)/r \rightarrow 0$, gives vanishing $\beta$ value. The value of anisotropic parameter and model parameter respectively as $k=0.95$ and $\alpha=3$ with $\beta=-0.05, -0.10, -0.15$. NEC$_r$ violates the null energy condition whereas NEC$_t$ satisfies the null energy condition. The difference between tangential and radial pressures $p_t-p_r$ decreasing from higher positive value and vanishes with the increase in the value of radial coordinate. The graphical behaviour of the energy condition $\rho+p_r+2p_t$ completely violating the energy condition and vanishing with increasing value of $r$.  \\

A comparative study between the previous model~\cite{Mishra21} and the present model, as far as NEC is concerned, we note that in the former model altogether the energy condition do violate for all the three cases. However, the present study shows that the situation is not so straight forward. In all three cases NEC violation does occur in the radial direction whereas in the tangential pressure direction fails to violate. \\

From the present investigation under $f(R)$ gravity it is explicitly clear that for construction of physically viable wormhole models we have not at all invoked any exotic matter, the so called dark matter. This result is in concordance of the other works on modified theories of gravities where the galactic dynamics of massive test particles has been explained without introducing any exotic dark energy~\cite{Carroll2004,Nojiri2007b,Nojiri2008,Cognola2008,Nojiri2011,Elizalde2011,Deb2018}. However, we would also like to highlight some literature survey which indicate the fact that - is exotic matter indispensable for constructing a traversable wormhole construction is indispensable with dark matter profile based on rotation curves~\cite{kuhfittig2014,rahaman2014a,rahaman2014b,rahaman2016a,rahaman2016b,Xu2020}. These two aspects are contradictory to each other and raise a serious question mark with the concrete validity of either theory, i.e., GR or modified gravity theories. It therefore can be theories that GR allows exotic entity for constructing wormhole whereas no need of an exotic component to form a stable wormhole structure. However, further extensive studies are essential before coming to a conclusive decision. \\

As a final remark, we say that, as matter of consequence of the topological censorship involving the non-violation of the null energy condition, traversable wormholes are not possible classically. However, quantum fields may provide enough negative energy for some wormholes to be traversable, for example we may note the asymptotically anti-de Sitter wormholes from gauge/gravity duality \cite{Gao2017, Maldacena2017, Bak2019}. Also, it has been shown that, traversable wormholes may be created through a quantum tunnelling effect by creating the wormhole mouths a finite distance apart with arbitrarily small acceleration \cite{Horowitz2019}. However, it is not yet clear about the class of spacetimes  allowing the creation of traversable wormholes through quantum mechanical route. Further, it is also not certain, whether the traversable wormhole could exist for long enough for any thing to pass through.
 
\section*{Acknowledgement} 
BM, SKT  and SR thank IUCAA, Pune, India for providing support through visiting associateship program. ASA acknowledges the financial support provided by University Grants Commission (UGC) through Senior Research Fellowship (File No. 16-9 (June 2017)/2018 (NET/CSIR)), to carry out the research work. The authors are thankful to the anonymous reviewer for the comments and suggestions to improve the quality of the paper.

\end{document}